**Reconfigurable chiral superconductivity**

Surajit Dutta[1†], Nadav Auerbach[1†], Tonghang Han[2†], Yaozhang Zhou[1], Gal Shavit[3],
Niladri-Sekhar Kander[1], Yuri Myasoedov[1], Martin E. Huber[4], Kenji Watanabe[5], Takashi Taniguchi[6],
Long Ju[2#], and Eli Zeldov[1*]

Rhombohedral multilayer graphene at high displacement fields hosts superconductivity emerging from a spin–valley-polarized quarter-metal, with transport signatures suggestive of time-reversal symmetry (TRS) breaking and chiral superconductivity (CSC). These observations have motivated proposals of topological superconductivity and non-Abelian quasiparticles, yet direct magnetic evidence and microscopic insight into the superconducting state remain lacking, limiting understanding of this unique state. Here we use nanoscale SQUID-on-tip magnetometry to image isospin-polarized domains in rhombohedral pentalayer graphene and establish CSC via spatially resolved thermodynamic detection of TRS breaking. We find that the density at which domain walls proliferate at elevated temperatures coincides with the onset of CSC, indicating an underlying transition in the parent state that both induces superconductivity and reduces domain-wall energy. We further show that the chiral domain structure in the superconducting phase is inherited from the isospin-polarized parent state. Strikingly, the CSC phase exhibits multiple transport regimes governed by configurations of chiral domains separated by highly resistive domain walls. We demonstrate deterministic, ultra-low-current control of these domains, enabling reversible switching between states of opposite chirality—a defining CSC property absent in other superconductors. These results establish rhombohedral graphene as a unique platform for reconfigurable CSC and ultra-low-power electronic functionality based on controllable isospin textures.

---

[1]Department of Condensed Matter Physics, Weizmann Institute of Science, Rehovot 7610001, Israel
[2]Department of Physics, Massachusetts Institute of Technology, Cambridge, MA, USA
[3]Department of Physics and Institute for Quantum Information and Matter, and Walter Burke Institute of Theoretical Physics, California Institute of Technology, Pasadena, California 91125, USA
[4]Departments of Physics and Electrical Engineering, University of Colorado Denver; Denver, Colorado 80217, USA
[5]Research Center for Electronic and Optical Materials, National Institute for Materials Science; 1-1 Namiki, Tsukuba 305-0044, Japan
[6]Research Center for Materials Nanoarchitectonics, National Institute for Materials Science; 1-1 Namiki, Tsukuba 305-0044, Japan
[†]These authors contributed equally to this work
[#]longju@mit.edu
*eli.zeldov@weizmann.ac.il

The emergence of unconventional superconductivity (SC) in strongly interacting electronic systems often coincides with symmetry breaking in the underlying state, giving rise to intertwined orders. In particular, superconductors with orbital TRS breaking can host chiral order parameters with finite angular momentum, offering a platform for non-abelian quasiparticles and Majorana-based quantum computing [1]. Rhombohedral multilayer graphene (RMG), with its topological flat bands and strong interactions, provides a clean and highly tunable platform for spin- and valley-polarized metals, integer and fractional Chern insulators, Wigner crystals, and superconductivity [2–17]. Notably, its two-dimensional nature and weak spin–orbit coupling enable a clear distinction between spin and valley (orbital) contributions to TRS breaking, allowing unambiguous identification of CSC without conflating it with spin-dominated effects. In this system, at large displacement fields, SC develops from a spin–valley-polarized quarter-metal (1/4M) phase and is accompanied by pronounced magnetic hysteresis and anomalous transport [8,14]. These observations raise two central and closely related unresolved questions: whether the SC phase is chiral, exhibiting macroscopic orbital TRS-breaking, and whether this chirality originates from the superconducting pairing symmetry itself or is inherited from the underlying correlated normal state [18,19]. In this work, we use the term CSC to denote superconducting states exhibiting macroscopic TRS breaking of orbital origin, irrespective of its microscopic mechanism.

Experimental signatures consistent with orbital TRS-breaking—most notably magnetic hysteresis and anomalous transport—have been reported in RMG [8,14] and interpreted as due to possible valley-polarized domains and switching. These experiments have motivated extensive theoretical proposals of unconventional and topological superconductivity [20–31]. However, transport measurements probe only global properties and do not resolve the real-space structure or thermodynamic nature of the ordered state, leaving the direct detection of magnetism and TRS-breaking in the SC phase unexplored. In RMG, the relevant internal degree of freedom is the spin–valley isospin associated with the fourfold manifold of electronic flavors. In the 1/4M phase, interactions lift this degeneracy and, in the presence of spin–orbit coupling [3,16,32], the ground state reduces to two nearly degenerate isospin polarizations, $K\uparrow$ and $K'\downarrow$, while the remaining two lie at higher energy. The system is therefore expected to support domains of opposite isospin polarization, with domain formation involving switching between these composite states. Understanding how isospin domains evolve as superconductivity emerges from the 1/4M phase is therefore a central question.

Domain walls (DWs) provide a powerful probe of intertwined phases, encoding both the energetics and mutual coupling. In conventional magnetic materials, DW structure and dynamics have been extensively studied and form the basis of spintronic technologies [33,34]. In contrast, their counterparts in isospin-polarized and strongly correlated two-dimensional systems [35–37] remain largely unexplored. In RMG, transport measurements have suggested that sharp resistance spikes and hysteresis arise from the formation of DWs between oppositely polarized states [14]. The domain structure has further been linked to the anomalous behavior of the fractional quantum anomalous Hall (FQAH) and the emergence of the extended quantum anomalous Hall effects in these materials [7,38,39]. However, the existence and dynamics of such domains have not been directly visualized, particularly in the SC regime. Establishing this structure across the superconducting transition offers a direct and unique route to probing the interplay between isospin polarization and SC.

Here, we use nanoscale SQUID-on-tip magnetometry to directly image and control isospin-polarized domains in rhombohedral pentalayer graphene (R5G). By combining local magnetic imaging with current-driven excitation, we access both the static configurations and real-space dynamics of DWs. We observe sharply defined domains with opposite out-of-plane magnetization, separated by highly resistive DWs, and track their evolution across the phase diagram. These domains persist deep within the SC phase, providing direct real-space thermodynamic evidence of TRS-breaking chiral domains— an experimental signature long sought but thus far absent in candidate chiral superconductors [40,41]. This establishes CSC with macroscopic TRS breaking in rhombohedral graphene. We further show that this TRS-breaking state is inherited from the underlying isospin-polarized phase. Beyond imaging, we demonstrate deterministic, ultra-low-current control of DW configurations, enabling reversible electrical switching at current scales orders of magnitude smaller than those required in spintronic systems [34]. The observed dynamics are consistent with a mechanism dominated by momentum transfer from reflected electrons, rather than spin-transfer torque, pointing to a new regime of low-dissipation control in isospin-polarized systems with potential applications in information technologies.

**Chiral superconductivity**

A hallmark of CSC is macroscopic TRS breaking, producing a finite intrinsic orbital magnetization. The sign of this magnetization is determined by which of the two degenerate macroscopic states, related by TRS, are spontaneously chosen by the system. To probe these signatures, we first characterize the transport properties of our devices.

Two R5G devices, intentionally misaligned from hBN, were fabricated with dual graphite gates (Figs. 1a,b and Methods). The carrier density $n$ and the displacement field $D$ were independently controlled via *dc* voltages $V_{tg}^{dc}$ and $V_{bg}^{dc}$ to the top and bottom gates, respectively (Fig. 1c). Owing to the non-standard contact geometry, we define the measured resistances as $R_{xx} \equiv V_{xx}/I$ and $R_{xy} \equiv V_{xy}/I$, where $I$ is the applied current (Figs. 1a,b). Figures 1d,e show $R_{xx}(n, D)$ and $R_{xy}(n, D)$ for device D1 in zero out-of-plane magnetic field ($B = 0$ T). The phase diagram comprises four regimes consistent with previous reports [3,14,16]: a half-metal (1/2M) at high carrier density, a quarter metal (1/4M) at intermediate density, a superconducting phase, and an insulating phase at the lowest densities (Methods and Extended Data Figs. 1,2). Figure 1g shows $R_{xx}(n)$ across the superconducting region at $B = 0$ mT and $D = -0.935$ V/nm (yellow dashed line in Figs. 1d,e) for different temperatures, revealing zero-resistance (ZR) state at our base temperature $T = 20$ mK, with a critical temperature $T_c \approx 150$ mK, defined using a 15% criterion of the normal-state resistance $R_{xx}$ (at $T = 560$ mK). This value of $T_c$ is consistent with the previously reported values extracted using BKT analysis in similar samples [14].

To directly visualize the magnetic texture, we performed nanoscale magnetic imaging using an indium SQUID-on-tip (SOT) magnetometer [42] (Fig. 1f and Methods). Optimal SOT sensitivity requires the application of a small magnetic field. The inset to Fig. 1g shows that the ZR state persists over the field range relevant for the SOT imaging presented here, $B < 40$ mT (see Extended Data Fig. 1a and 2b,e for the full field dependence). To probe the magnetic state, a small *ac* voltage modulation $V_{bg}^{ac}$ or $V_{tg}^{ac}$ is superimposed on *dc* backgate $V_{bg}^{dc}$ or topgate $V_{tg}^{dc}$ voltages (Fig. 1c), inducing a corresponding oscillation in the carrier density $n^{ac}$. This modulates the local out-of-plane magnetization $M_z$, generating an oscillatory stray field $B_z^{ac}$ detected by the SOT (Fig. 1f). The measured signal $B_z^{ac} \propto n^{ac} dM_z/dn$ thus provides a direct local probe of the differential magnetization $m_z = dM_z/dn$ [16,42], reflecting the magnetization carried by the electrons that are periodically added to and removed from the system.

Figure 1h shows $B_z^{ac}$ map acquired in device D1 at $T = 20$ mK and $B = 26$ mT within the CSC phase in the ZR state (green circle in Figs. 1d,e). In contrast to the expected diamagnetic response of a superconductor, the positive $B_z^{ac}$ signal establishes a ferromagnetic (or paramagnetic) response. Repeating the measurement at $B = -28$ mT yields a negative $B_z^{ac}$ (Fig. 1i), again consistent with ferromagnetism. If the TRS breaking originates from spontaneous orbital chirality, the system should be trainable between opposite chiral states. This is demonstrated in Figs. 1j,k, both acquired at the same field $B = 20$ mT. Initializing the system at $B_{int} = 120$ mT produces a positive $B_z^{ac}$ (Fig. 1j), whereas initialization at $B_{int} = -100$ mT reverses the signal to negative $B_z^{ac}$ under otherwise identical conditions (Fig. 1k). Together, these results provide direct, spatially resolved evidence of TRS breaking, revealing two degenerate superconducting states of opposite chirality. This constitutes unambiguous thermodynamic evidence for CSC in RMG.

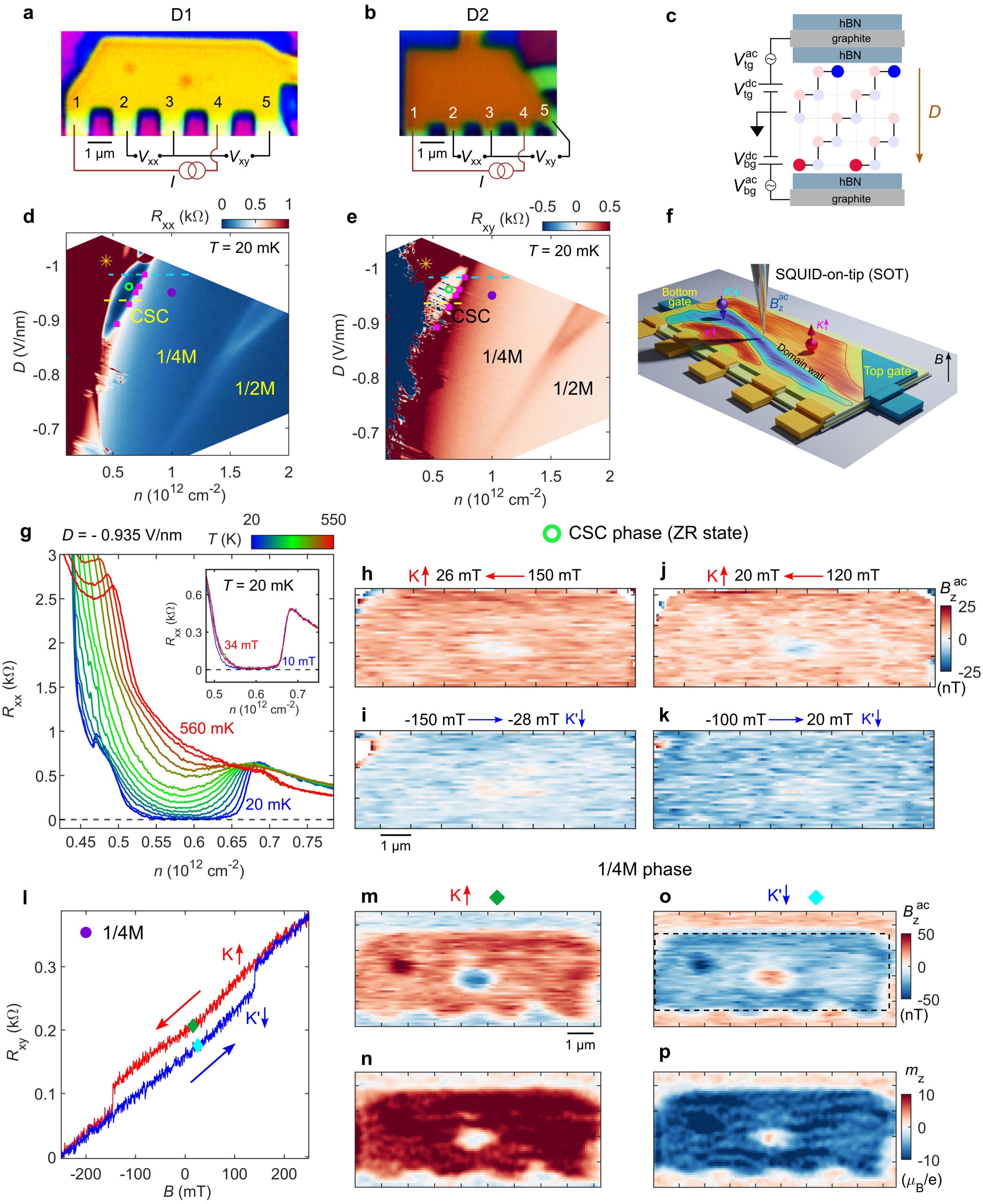

**Fig. 1. Magnetic response of CSC and 1/4M phases. a,b,** Optical images of devices D1 (**a**) and D2 (**b**) showing electrical contacts labeled 1 to 5 used for transport measurements, with $V_{xx} \equiv V_3 - V_2$, $R_{xx} \equiv V_{xx}/I$, $V_{xy} \equiv V_5 - V_3$, and $R_{xy} \equiv V_{xy}/I$. The two darker circular regions in device D1 correspond to bubbles. **c,** Schematics of the R5G device and gate-voltage circuitry. **d,** Longitudinal resistance $R_{xx}(n, D)$ measured at $B = 0$ mT and $T = 20$ mK, showing the CSC, 1/4M, and 1/2M phases in device D1. Magenta squares mark the onset density of spike-like hysteresis $n_{DW}$ (Extended Data Fig. 1). **e,** Same as (d), for Hall resistance $R_{xy}(n, D)$. **f,** Schematic of the SOT measurement configuration. **g,** Temperature dependence of $R_{xx}$ measured along the yellow dashed line in (d,e) showing the ZR state

at low temperatures in the CSC phase. The inset shows the magnetic-field dependance of $R_{xx}$ at $T =$ 20 mK. Four nearly overlapping curves for $B = 10$, 18, 26, and 34 mT indicate minimal field dependence over the range used for SOT imaging ($B < 40$ mT) in the ZR state. **h,** $B_z^{ac}$ map measured in the CSC phase ($D = -0.961$ V/nm, $n = 0.63\times10^{12}$ cm$^{-2}$, green circle in (d,e)) in the ZR state using backgate modulation $V_{bg}^{ac}$ at $B = 26$ mT, following initialization at positive $B_{int} = 150$ mT, yielding positive magnetization chirality $K\uparrow$. **i,** Same as (h), measured at negative $B = -28$ mT following initialization at negative $B = -150$ mT, yielding negative chirality $K'\downarrow$. **j,** $B_z^{ac}$ map measured at positive $B = 20$ mT following initialization at $B_{int} = 120$ mT, yielding positive chirality $K\uparrow$. **k,** Same as (j), measured at the same $B = 20$ mT but following initialization at negative $B_{int} = -100$ mT, yielding negative chirality $K'\downarrow$. **l,** Magnetic hysteresis of $R_{xy}$, measured in the 1/4M phase at high carrier density ($D = -0.95$ V/nm, $n = 1\times10^{12}$ cm$^{-2}$, purple dot in (d,e)). **m,** $B_z^{ac}$ map acquired in the 1/4M phase at the upper branch of the hysteresis loop (green diamond in (l)). The two circular features in $B_z^{ac}$ correspond to bubble locations visible in (a). **n,** Corresponding reconstructed map of the differential magnetization $m_z$, revealing a spatially uniform positive magnetization, consistent with $K\uparrow$ isospin polarization. **o,p** same as (m,n) but measured on the lower branch of the hysteresis loop (cyan diamond in (l)), showing spatially homogeneous negative magnetization corresponding to $K'\downarrow$ isospin polarization. The dashed frame in (o) marks the experimental scan area in (h-k).

We next show that the chirality in the CSC phase is inherited from the parent 1/4M phase. The spin–valley-polarized 1/4M phase spontaneously breaks orbital TRS, manifested by anomalous Hall effect (AHE) [3,14,16]. At high carrier densities, the magnetic hysteresis exhibits a conventional box-like loop characteristic of a ferromagnet, with a characteristic coercive field $B_{co} \approx \pm150$ mT (Fig. 1l). To determine the corresponding magnetic texture, we acquire $B_z^{ac}$ map on the upper branch of the hysteresis loop at $B = 16$ mT (green diamond in Fig. 1l), following $B_{int} = 250$ mT. Figure 1m shows that under these conditions, the sample exhibits a nearly uniform positive $B_z^{ac}$ signal across the device, aside from two small bubble locations visible in the optical image in Fig. 1a. By performing a numerical inversion [16,43], we reconstruct the local differential magnetization $m_z$, which reveals a rather uniform magnetization with large average value of 9.5 $\mu_B$ per electron (Fig. 1n), indicating the dominance of Berry-curvature–induced orbital magnetization [16,42]. Since in the 1/4M phase spin–orbit coupling locks the spin and valley moments [16,44], the extracted $m_z$ reflects the total magnetization associated with the spin–valley isospin order parameter. The positive $m_z$ map therefore indicates a homogeneous state polarized to the $K\uparrow$ isospin. Notably, this large magnetization is characteristic of the 1/4M phase at high carrier densities and decreases strongly with decreasing $n$ (Extended Data Fig. 2d), accounting for the much weaker $B_z^{ac}$ signal observed in Figs. 1h-k.

Figure 1o shows the $B_z^{ac}$ map acquired on the lower branch of the hysteresis loop at $B = 25.5$ mT (cyan diamond in Fig. 1l), following $B_{int} = -250$ mT. The signal exhibits a uniform polarity opposite to that of the upper branch. The corresponding $m_z$ map reveals a uniform negative magnetization (Fig. 1p), indicating homogeneous polarization to the $K'\downarrow$ isospin. Thus, in the 1/4M phase at elevated doping, each branch of the box-like hysteresis loop corresponds to a spatially uniform isospin-polarized state, with no coexistence of oppositely polarized domains. Together, these results show that TRS breaking is intrinsic to the parent 1/4M phase and is directly inherited by the CSC state.

**Transport hysteresis in CSC and 1/4M phases**

Strikingly, in addition to the ZR state shown in Figs. 1g-k, the CSC phase can exhibit a rich and tunable transport phenomenology. Because CSC supports two degenerate chiral states, their coexistence naturally gives rise to DWs separating regions of opposite chirality. We next show that these domains can be controlled by both magnetic field and applied current, enabling a reconfigurable CSC state. This control produces distinct transport regimes within the CSC phase, including zero-resistance (ZR), low-resistance (LR), high-resistance (HR), and even negative-resistance (NR) states (Methods).

We begin with the HR state. Figures 2a,b show the magnetic hysteresis of $R_{xy}$ and $R_{xx}$ at $T = 20$ mK in the high-density 1/4M phase, exhibiting conventional box-like loops similar to Fig. 1l. Strikingly, upon entering the CSC phase at lower $n$, this behavior changes abruptly: a pronounced spike-like hysteresis emerges reaching several kΩ (green and brown traces in Figs. 2e,f), thereby forming a HR state from an otherwise ZR state. This behavior is consistent with previous reports [14]. The detailed characteristics of the HR state depend on parameters including $n$, $D$, sweep rates, and magnetic and thermal history (Extended Data Fig. 3 and Methods).

To quantify this evolution, we analyze the hysteresis $\Delta R_{xy}(B,n) = R_{xy}(B\uparrow,n) - R_{xy}(B\downarrow,n)$ and $\Delta R_{xx}(B,n) = R_{xx}(B\uparrow,n) - R_{xx}(B\downarrow,n)$ as a function of $n$ at fixed $D$ (Figs. 2i,j and Extended Data Figs. 1d-k). At high $n$, the box-like hysteresis yields a uniform negative $\Delta R_{xy}$ (light blue in Fig. 2i) within $|B| < |B_{co}|$, consistent with a uniform reversal of the isospin-polarized state. However, below a characteristic density, $n_{DW} = 0.77 \times 10^{12}$ cm$^{-2}$ (magenta dashed lines in Figs. 2i,j and top magenta mark in Figs. 1d,e), the box-like loop is replaced by sharp, spike-like features characteristic of the HR state. Remarkably, across different $D$, the onset density $n_{DW}(D)$ of the spike-like features closely tracks the high-density boundary of the CSC region (magenta marks in Figs. 1d,e and Extended Data Fig. 1). This correspondence is also reproduced in device D2 (Extended Data Fig. 2). These observations may initially suggest that the onset of the resistance spikes coincides with the emergence of the CSC order parameter.

However, Figs. 2c-l show that the same abrupt transformation to spike-like hysteresis and HR state occurs at $T = 1$ K in the normal phase, well above $T_c$ (Extended Data Figs. 1l,m). Moreover, the transformation occurs at the same $n_{DW}(D)$ in both the CSC and 1/4M phases. This raises a central question: why does the spike-like transformation emerge in the 1/4M phase at elevated temperature precisely at the density where CSC develops at lower temperatures? Such systematic correspondence strongly points to a common underlying mechanism.

In addition to the HR state, we often observe an unusual LR state with finite $R_{xx}$ and $R_{xy}$ in CSC phase, which also exhibits a small but clear hysteresis upon sweeping magnetic field (blue and red traces in Figs. 2e,f and corresponding blue and red backgrounds in the CSC phase in Figs. 2i,j and Extended Data Figs. 1,2). Notably, as discussed below, thermal cycling can remove the LR state, restoring the ZR state.

**Imaging isospin domains in the 1/4M phase**

To elucidate the origin of the HR state and enable direct comparison between the 1/4M and CSC phases, we first perform SOT magnetic imaging in 1/4M phase at base temperature ($T = 20$ mK). At this temperature and for $n < n_{DW}$, the HR state can be accessed within the 1/4M phase—rather than the CSC phase—by applying a larger displacement field $|D|$, beyond the CSC regime (yellow asterisk in Figs. 1d,e). In this regime, the system can occupy either the LR or HR state, depending on the magnetic-field sweep history. In the LR state (black star in Figs. 2m,n), the $B_z^{ac}$ image displays a uniform positive signal (red in Fig. 2o), indicating homogeneous $K\uparrow$ isospin polarization across the sample, which we refer to as the majority polarization. In contrast, in the HR state (blue star in Figs. 2m,n) the bottom-left region exhibits a negative (blue) $B_z^{ac}$ (Fig. 2p), revealing a minority domain polarized to $K'\downarrow$. This contrast is highlighted in Fig. 2q, which shows the numerical subtraction of the LR and HR images. The top-right region shows near-zero signal (white), indicating unchanged polarization $K\uparrow$, while the pronounced red signal in the bottom-left reflects a reversal to minority polarization. These measurements show that the LR state corresponds to a nearly uniform isospin polarization, whereas the HR state hosts a $K\uparrow$-$K'\downarrow$ DW traversing the sample (green dotted line in Fig. 2q).

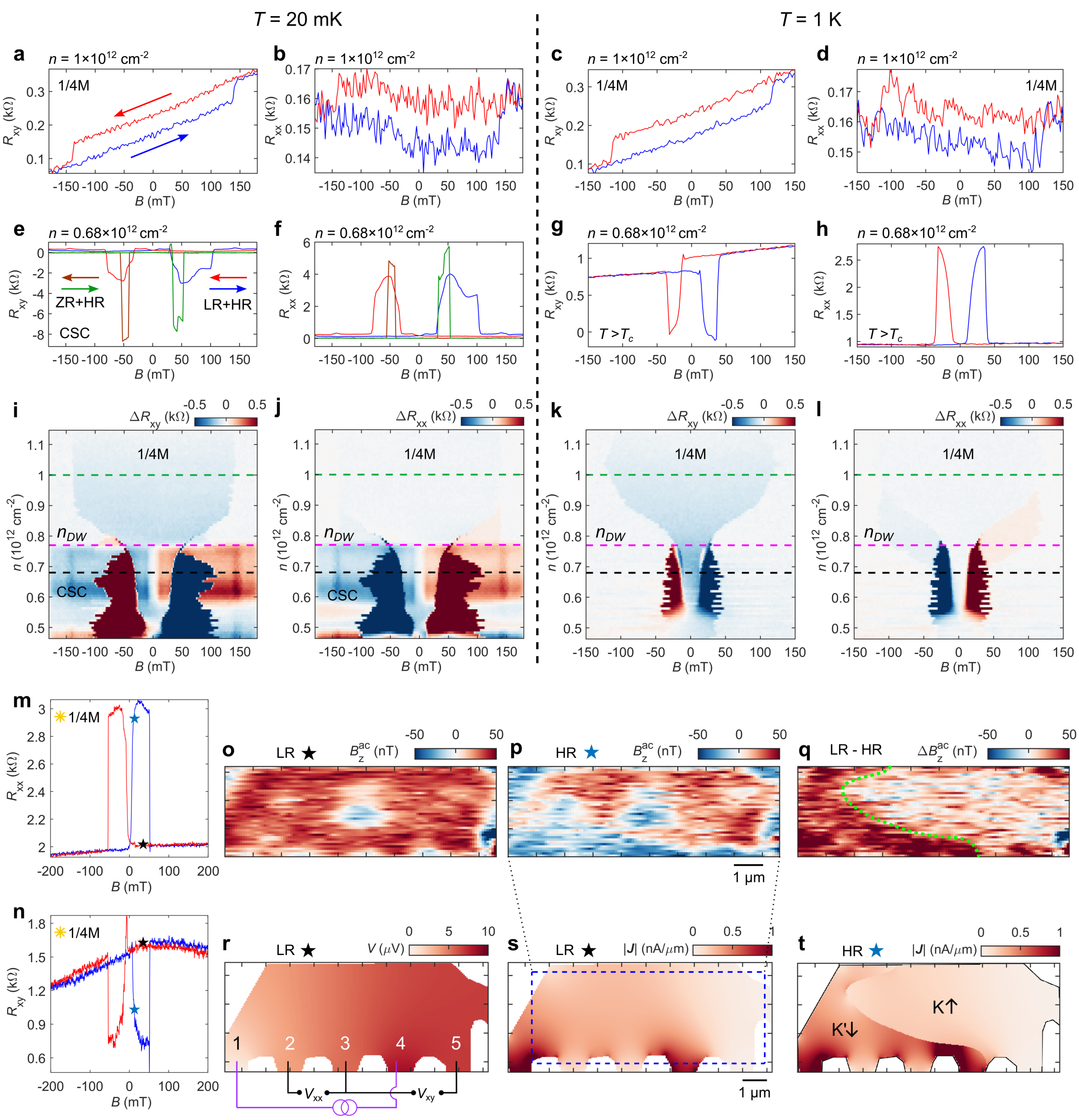

**Fig. 2. Transport hysteresis and imaging of isospin-polarized domains in the 1/4M phase. a,b,** Magnetic field hysteresis of $R_{xy}$ (**a**) and $R_{xx}$ (**b**) in device D1 in the 1/4M phase ($D = -0.983$ V/nm, $n = 1\times10^{12}$ cm$^{-2}$) along the green dashed line in (i,j) at $T = 20$ mK, showing box-like hysteresis. **c,d,** Same as (a,b) measured at $T = 1$ K. **e,f,** Same as (a,b) measured at $n = 0.68\times10^{12}$ cm$^{-2}$ $< n_{DW}(D)$ in the CSC phase along the black dashed line in (i,j), showing low resistances in the LR state and large resistance spikes in the HR state (blue and red LR+HR curves). After thermal cycling (Methods), a similar measurement at $D = -0.957$ V/nm and $n = 0.65\times10^{12}$ cm$^{-2}$ shows vanishing resistances in ZR state (green and brown ZR+HR curves). **g,h,** Same as (e,f) measured at $T = 1$ K $> T_c$, showing similar resistance spikes in the parent state. **i,** Magnetic-field hysteresis $\Delta R_{xy}(B,n) = R_{xy}(B\uparrow,n) - R_{xy}(B\downarrow,n)$, measured along the cyan dashed line in Figs. 1d,e at $D = -0.983$ V/nm and $T = 20$ mK. $R_{xy}(B\uparrow,n)$ and $R_{xy}(B\downarrow,n)$ were measured by sweeping $B$ from $-0.2$ T to 0.2 T and back (blue and red traces in (a)). The magenta dashed line marks the onset density $n_{DW}(D)$ of spike-like hysteresis due to DW proliferation both above and below $T_c$ (magenta marks in Figs. 1d,e). **j,** Corresponding measurement of $\Delta R_{xx}(B,n) = R_{xx}(B\uparrow,n) - R_{xx}(B\downarrow,n)$. **k,l,** Same as (i,j), measured at $T = 1$ K $> T_c$, showing similar behavior in the parent state. **m,n,** Magnetic hysteresis of $R_{xx}$ (**m**) and $R_{xy}$ (**n**), showing

pronounced HR-state spikes in the high-$D$ 1/4M phase ($D = -1.008$ V/nm, $n = 0.44\times10^{12}$ cm$^{-2}$, yellow asterisk in Figs. 1d,e) at $T = 20$ mK, similar to the CSC phase in (e,f). **o,** Spatial map of $B_z^{ac}$ in the LR state (black star in (m,n)), showing uniform positive magnetization. **p,** Same as (o) in the HR state (blue star in (m,n)), revealing a minority domain with negative magnetization in the lower-left region. **q,** Difference map between $B_z^{ac}$ maps in LR (o) and HR (p) states, highlighting the DW (green dotted line). **r,** Numerical simulation of the potential map $V$ in the homogeneous LR state with $I = 1$ nA applied between contacts 4 and 1, using $\sigma_{xx}$ and $\sigma_{xy}$ fitted to $R_{xx}$ and $R_{xy}$ (black star in (m,n)), along with the corresponding circuit schematic. **s,** Corresponding map of the current density magnitude $|\boldsymbol{J}|$. The blue frame marks the experimental scan area in (o-p). **t,** Same as (s) for the HR state, incorporating a DW positioned as in the experiment (q), with fitted wall resistivity $\rho_{DW} = 41$ kΩ·µm, reproducing the measured $R_{xx}$ and $R_{xy}$ (blue star in (m,n)).

The striking contrast of several kΩ between the LR and HR states (Figs. 2m,n) indicates that the isospin DW introduces a large additional resistance, implying a substantial 1D interfacial resistivity $\rho_{DW}$ across the boundary. To quantify $\rho_{DW}$, we model the current flow in the device by numerically solving the 2D transport equations in the experimental geometry (Fig. 1a), using COMSOL simulations. In the LR state we assume spatially homogeneous longitudinal and transverse conductivities, $\sigma_{xx}$ and $\sigma_{xy}$, and determine their values by fitting the calculated $R_{xx}$ and $R_{xy}$ to the measured values (black star in Figs. 2m,q), yielding corresponding bulk resistivities $\rho_{xx} = 2.57$ kΩ and $\rho_{xy}$ =281.5 Ω. The resulting simulated potential and current-density distributions are shown in Figs. 2r,s.

For the HR state, we retain the same bulk conductivities $\sigma_{xx}$ and $\sigma_{xy}$, but introduce a 1D DW along the experimentally observed trajectory with an interfacial resistivity $\rho_{DW}$. We additionally reverse the sign of $\sigma_{xy}$ in the minority $K'\downarrow$ domain to account for the opposite isospin polarization. Treating $\rho_{DW}$ as a fitting parameter to reproduce the measured $R_{xy}$ and $R_{xx}$ in the HR state (blue star in Figs. 2m,n) yields $\rho_{DW} = 41$ kΩ·µm. The corresponding simulated current-density map (Fig. 2t) shows that most of the current between contacts 4 and 1 is confined to the minority domain spanning both contacts, with only a small fraction crossing the DW into the majority domain due to the large $\rho_{DW}$. In this model, we assume a uniform $\rho_{DW}$ along the DW for simplicity. Microscopically, however, $\rho_{DW}$ may vary spatially depending on the orientation of the DW relative to the crystallographic graphene axes. A detailed investigation of these effects is beyond the scope of the present work and is left for future studies.

This first experimental imaging and estimate of the isospin DW resistivity in RMG provides important insight into inter-domain charge transport. Its large value reflects a strongly suppressed electron tunnelling probability across the interface (Methods), as transmission requires both a large momentum transfer between opposite valleys and a spin flip associated with the reversal of spin polarization. This strong suppression renders the DW effectively opaque to charge transport, giving rise to the pronounced resistance spikes observed in the HR state.

**Imaging the LR and HR states in the CSC phase**

Figures 1h-k show that in the ZR state the sample is uniformly polarized, forming a single isospin-majority domain. We next visualize the isospin structure underlying the LR and HR states in the CSC phase. In the HR state (blue star in Figs. 3a,b), the $B_z^{ac}$ map in Fig. 3c reveals coexisting majority (red) and extended minority (blue) domains that span multiple contacts at the bottom of the sample, along with a few smaller domains in the bulk. Notably, contacts 2 and 3, across which $R_{xx}$ is measured, reside in different domains, separated by highly resistive DWs, thereby giving rise to the HR state, as discussed in more detail below. In contrast, the LR state (black star in Figs. 3a,b) contains only small, spatially confined minority domains that do not connect multiple contacts (Fig. 3d), resulting in a low-resistance state. These minority domains likely nucleate at edges or disorder sites in the bulk, where the barrier

for isospin reversal is reduced, but remain confined due to DW pinning. These observations establish that the transition between LR and HR states is governed by a geometric reconfiguration of domains from confined bubbles to system-spanning structures. They further show that the DW-dominated transport observed in the HR state—also present in the 1/4M phase at $T > T_c$ and at high $D$ (Figs. 2m-t)—persists across the superconducting transition and is directly inherited from the parent isospin-polarized state.

Moreover, the hysteretic behavior is stochastic and dependent on position in the phase diagram and magnetic-field sweep conditions (see Methods and Extended Data Fig. 3). This variability is illustrated in Figs. 3g,h, which show markedly different domain configurations for opposite $B$ polarities, leading to HR states that are strongly pronounced in $R_{xy}$ but exhibit only a weak signature in $R_{xx}$ (blue and purple stars in Figs. 3e,f). These diverse regimes and stochastic fluctuations are discussed in Methods and Extended Data Fig. 3. Notably, the metastable small domains responsible for the LR state can be annealed by heating the sample above 10 K followed by zero-field cooling, thereby reestablishing the ZR state (green and brown curves in Figs. 3a,b) with uniform isospin structure (Figs. 1h-k).

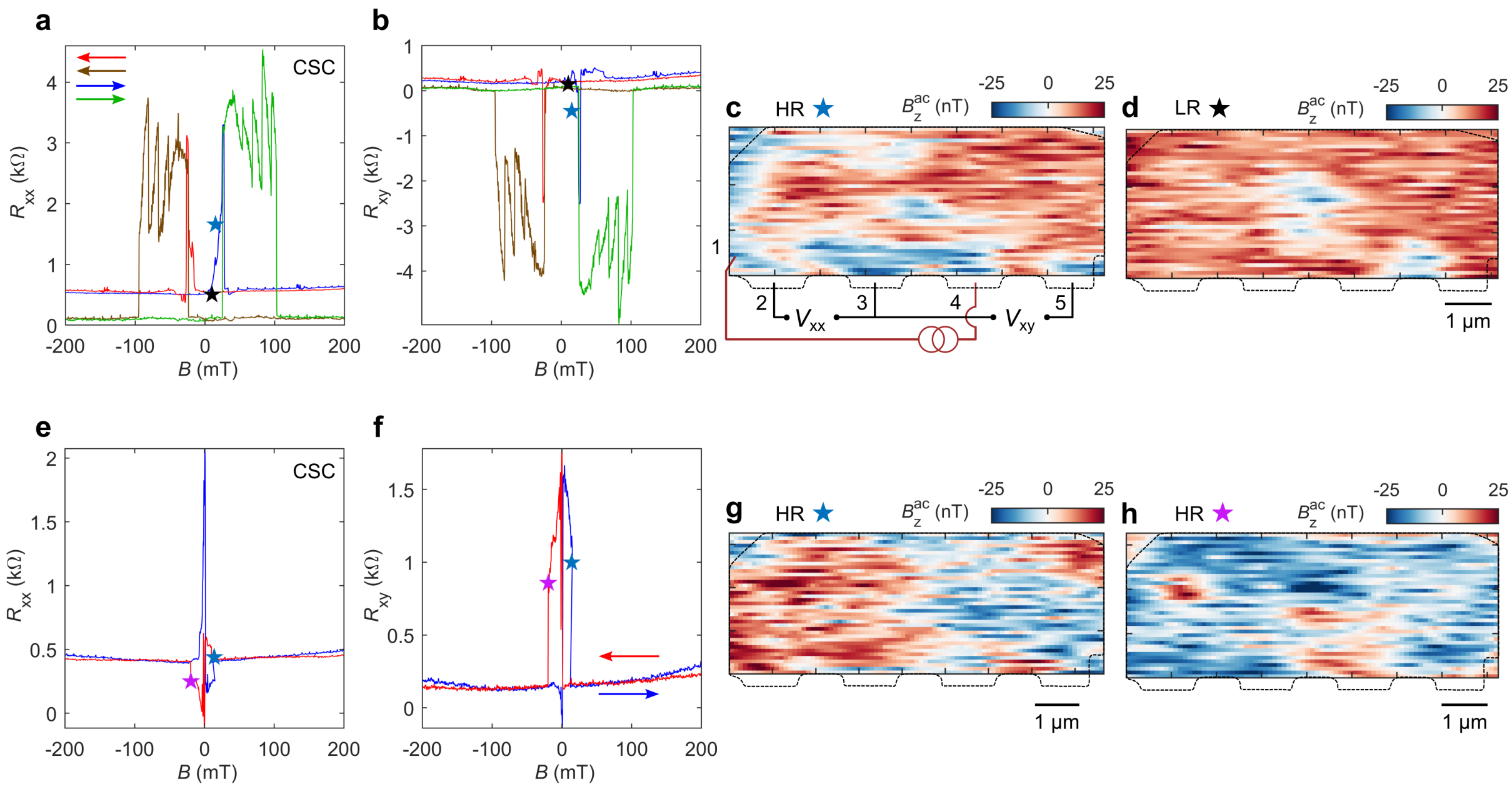

**Fig. 3. Imaging of isospin-polarized domains in the CSC phase. a,b,** Magnetic-field hysteresis of $R_{xx}$ (**a**) and $R_{xy}$ (**b**) in device D1 measured in the CSC phase ($D = -0.983$ V/nm, $n = 0.66\times10^{12}$ cm$^{-2}$). Red and blue curves were obtained after multiple field sweeps, showing LR+HR states, whereas brown and green curves were measured following thermal cycling, showing ZR+HR hysteretic behavior similar to Figs. 2e,f. Stars indicate the resistance values measured during imaging in (c,d) at fixed $B$. **c,** Spatial map of $B_z^{ac}$ in the HR state (blue star in (a,b)), revealing an extended minority domain (blue) spanning several contacts, yielding an HR configuration. The sample boundary is delineated by the black dotted line along with the electric circuit. **d,** Same as (c), measured in the LR state (black star in (a,b)), revealing small, spatially confined minority domains that do not connect multiple contacts, giving rise to an LR configuration, in contrast to the ZR state with no minority domains (Figs. 1h-k). **e,f,** Same as (a,b), measured at a different point in the CSC phase ($D = -0.977$ V/nm, $n = 0.66\times10^{12}$ cm$^{-2}$). **g,h,** $B_z^{ac}$ maps measured in the HR state (blue and purple stars in (e,f)) for positive (**g**) and negative (**h**) magnetic fields, showing extended minority domains.

We now return to the question of why DW proliferation in the 1/4M phase at elevated temperatures coincides with the onset density of CSC at low temperatures. We propose that the 1/4M undergoes an additional hidden symmetry-breaking transition at $n_{DW}(D)$ (Methods), introducing a secondary order parameter with two key effects: (i) coinciding with the onset of superconductivity, suggesting a reconstructed parent state favorable for superconductivity, and (ii) reducing the DW line tension (Extended Data Fig. 11), thereby facilitating nucleation of opposite-polarity domains and proliferation of DWs. As a precursor to this transition, the coercive field $B_{co}$ decreases rapidly as $n$ approaches $n_{DW}$ from higher values, a behavior observed both above and below $T_c$ (Figs. 2i-l). In addition, the spike-like hysteresis below $n_{DW}$ also appears when sweeping $n$ instead of $B$ (Extended Data Figs. 1n,o). Taken together, these observations point to an intimate connection between superconductivity and an additional, yet unidentified symmetry-breaking transition in the parent state that simultaneously enhances CSC and reduces the DW energy (Methods).

**Current-induced DW dynamics**

While the DW configurations discussed so far are primarily determined by the magnetic-field sweep history, we now show that they can also be generated and manipulated using an applied electric current. Application of a small current significantly narrows the box-like hysteresis (Extended Data Fig. 4), highlighting the close correspondence between field- and current-driven dynamics of isospin domains. We begin by examining current-driven configurations in the high-$D$ 1/4M phase at $T = 20$ mK at densities below the $n_{DW}(D)$ line. Figure 4a shows $B_z^{ac}$ map in device D2 acquired in the LR state in the absence of applied current ($I_{dc} = 0$ nA). The positive $B_z^{ac}$ signal indicates that the sample is predominantly polarized to majority $K\uparrow$ isospin. Applying $I_{dc} = 20$ nA produces no visible change in the magnetic texture (Fig. 4b). However, at $I_{dc} = 30$ nA a blue domain nucleates at contact 1, which serves as the electron-injection (source) contact, revealing the formation of minority $K'\downarrow$ domain (Fig. 4c). This minority bubble expands further as the current is increased to $I_{dc} = 40$ nA (Fig. 4d and Fig. 4h schematic), and collapses upon reducing the current back to zero (Fig. 4e).

Despite the presence of substantial stochastic fluctuations—visible as streaks in transport in Figs. 1d,e, noise in Extended Data Figs. 3v,w, and random domain reconfigurations in Extended Data Fig. 5—the expansion and contraction of the bubble with increasing and decreasing $I_{dc}$ are reproducible on average (Extended Data Fig. 6), and form a quasi-static configuration (Extended Data Fig. 7). Small minority bubbles can nucleate and annihilate spontaneously due to fluctuations and disorder, particularly along the sample edges. When such a bubble forms at the interface between a current contact and the RMG, its subsequent evolution is strongly influenced by the current direction: electron injection from the metal contact drives bubble expansion, whereas electron extraction leads to its contraction. This picture naturally accounts for the pronounced current-polarity asymmetry, as demonstrated in Fig. 4f, where no bubble nucleation is observed for negative current $I_{dc} = -45$ nA.

Since the DW dynamics are largely reversible, they can be probed using an *ac* current $I_{ac}$ instead of $I_{dc}$. This approach offers two advantages: (i) It eliminates the need for gate-voltage modulation to generate the $B_z^{ac}$ signal, as the DW displacement itself produces the *ac* response detected by the SOT. (ii) Under an *ac* drive, the DW behaves as an elastic string in a disorder-induced 2D random potential, oscillating at the drive frequency. As the wall moves back and forth, the regions it traverses switch between majority to minority isospin polarization, with the corresponding magnetization reversing from $M_z$ to $-M_z$ (and vice versa). Because all carriers within the swept region reverse their isospin, this results in a substantially larger local magnetization modulation, $\Delta M_z \approx 2M_z$, compared to the gate-induced differential response $m_z^{ac} \approx n^{ac} dM_z/dn$. In addition, regions outside the DW excursion range remain static and therefore exhibit vanishing $B_z^{ac}$, enhancing the contrast between oscillating and stationary areas.

This behavior is demonstrated in Fig. 4i, measured in the CSC state in device D1 at $B = -25$ mT. An *ac* current $I_{ac} = 5.3$ nA is applied, and the $B_z^{ac}$ map is measured at the drive frequency ($f = 1.181$ kHz). The blue region on the left reflects the periodic expansion and contraction of a minority-domain bubble driven by current injection at contact 1. During the positive half-cycle of $I_{ac}$, the injected electrons (at contact 1) nucleate and expand the $K'{\downarrow}$ bubble (Fig. 4j schematic), whereas during the negative half-cycle, when electrons are extracted at contact 1, the bubble collapses (Fig. 4k). The resulting $B_z^{ac}$ map reflects the difference between the configurations in Figs. 4j and 4k, shown schematically in Fig. 4l, with zero signal elsewhere in the sample where no switching occurs.

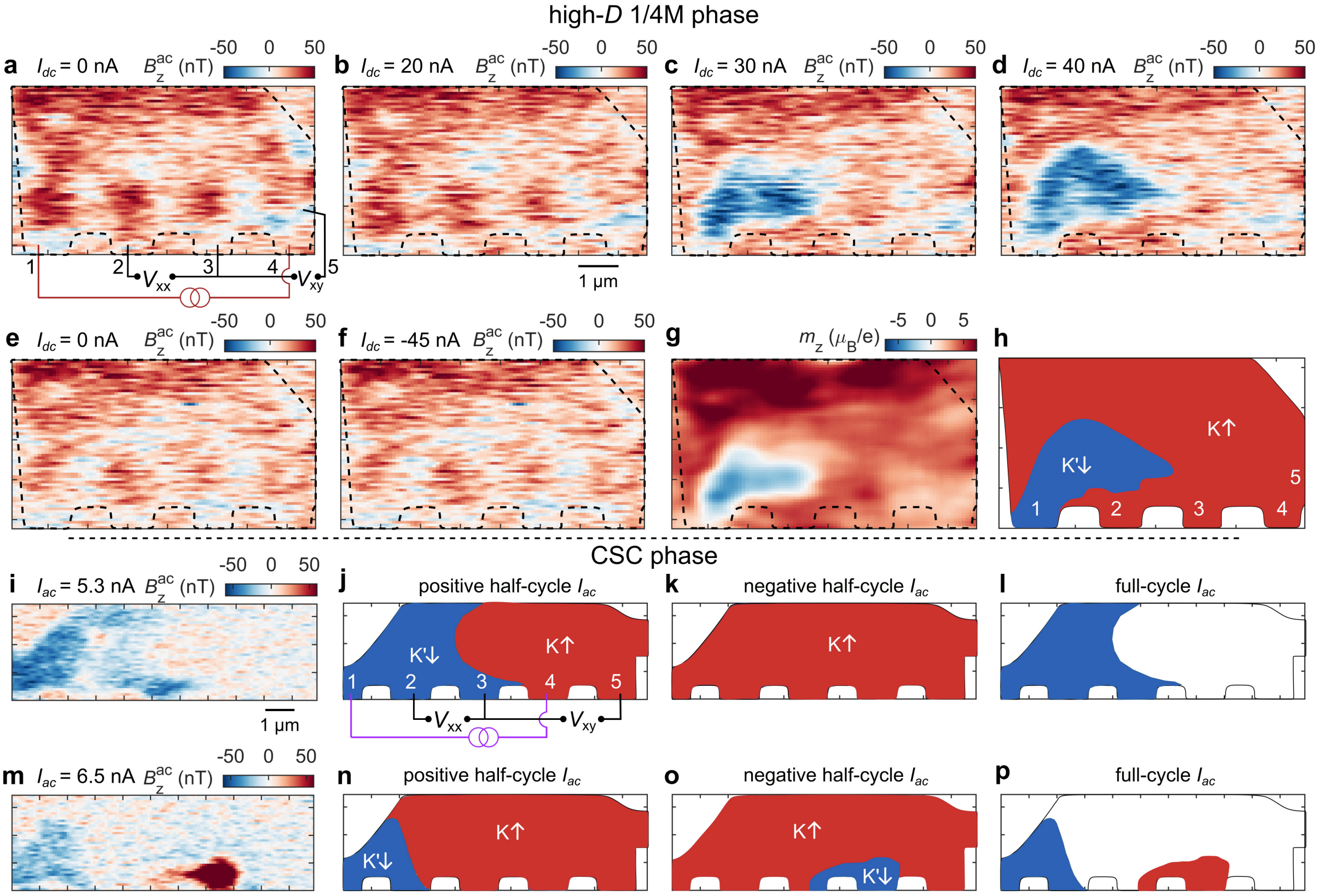

**Fig. 4. Current-induced dynamics of isospin domains in the 1/4M and CSC states. a-f,** $B_z^{ac}$ maps acquired using backgate modulation $V_{bg}^{ac}$ in the high-$D$ 1/4M phase ($D = -1.03$ V/nm, $n = 0.422\times10^{12}$ cm$^{-2}$, $T = 20$ mK) in device D2 under the indicated *dc* currents at $B = 13$ mT. Black dashed lines delineate the sample boundaries. **g,** Differential magnetization map $m(x, y)$ reconstructed from (c). **h,** Schematic isospin polarization map corresponding to (d), showing the formation of a minority $K'{\downarrow}$ domain near contact 1 due to electron injection. **i,** $B_z^{ac}$ map measured in the CSC phase in device D1 ($D = -0.996$ V/nm, $n = 0.75\times10^{12}$ cm$^{-2}$, $T = 20$ mK) at $B = -25$ mT using current modulation $I_{ac} =$ 5.3 nA at frequency $f = 1.181$ kHz, showing a unipolar domain formation (blue). **j,** Schematic domain structure corresponding to (i), where the blue region denotes a minority $K'{\downarrow}$ domain formed due to electron injection at contact 1 during the positive half-cycle of $I_{ac}$. **k,** Schematic of the uniformly polarized majority $K{\uparrow}$ state during the negative half-cycle of $I_{ac}$. **l,** Difference map between (j) and (k), corresponding to the $B_z^{ac}$ signal in (i). **m,** Same as (i) but at $B = -20$ mT, showing a bipolar configuration with two minority domains near contacts 1 and 4. **n,** Schematic domain structure corresponding to (m) with minority domain formed due to electron injection at contact 1 during the positive half-cycle of $I_{ac}$. **o,** Same as (n), but with minority domain formed due to electron injection at contact 4 during the negative half-cycle of $I_{ac}$. **p,** Difference map between (n) and (o), reproducing the experimental $B_z^{ac}$ signal in (m).

We refer to the behavior described above as a unipolar regime, in which the domain dynamics are confined to a single contact that acts as the source (drain) during the negative (positive) half-cycle of the *ac* drive. In contrast, Fig. 4m shows a bipolar regime at $B = -20$ mT and $I_{ac} = 6.5$ nA in the same CSC state. In this case, during the positive half-cycle a minority domain expands near contact 1, where electrons are injected (blue bubble and Fig. 4n schematic), whereas during the negative half-cycle a minority domain forms near contact 4, as electron injection shifts to that contact. As a result, the $B_z^{ac}$ map (Fig. 4m), which reflects the difference between configurations in Figs. 4n and 4o, exhibits opposite-polarity signals corresponding to magnetization oscillations that are out of phase near the two contacts (Fig. 4p). Remarkably, DW dynamics can be resolved at currents as low as 0.4 nA (Extended Data Fig. 8g), representing the lowest magnetic domain manipulation current reported to date [35,36,45]. This reversible switching between states of opposite chirality highlights a defining property of the CSC phase in RMG—macroscopic, switchable TRS breaking—absent in conventional superconductors.

Moreover, the extremely low currents required to drive DW dynamics—comparable to those typically used in transport measurements—imply that transport measurements in the CSC phase cannot be regarded as purely passive, linear-response probes of a static system. Instead, they can be invasive, inducing DW dynamics and the associated dissipation. This effect likely contributes to the finite resistance observed in the LR state and is at the root of the NR state (Methods) within the CSC phase.

**Current-driven resistance switching**

The combination of vanishing bulk resistivity in the CSC state and the large interfacial resistivity of isospin DWs provides a natural mechanism for ultra-low-current resistance switching, a key functionality in electronic and spintronic devices. To demonstrate this, we apply simultaneous *dc* and *ac* currents (Fig. 5e schematics), where $I_{dc}$ acts as a control parameter and a small probing current $I_{ac} = 1$ nA is used to measure the resistive state. We periodically switch $I_{dc}$ between $+1.7$ nA and $-1.7$ nA in the CSC state (Fig. 5a). At $I_{dc} = -1.7$ nA, the device resides in the LR state with $R_{xx} \cong 0\ \Omega$ (Fig. 5b) and $R_{xy} \cong 500\ \Omega$ (Fig. 5c). Strikingly, reversing the current to $I_{dc} = +1.7$ nA switches the system into the HR state, with $R_{xx} \approx 2.7$ kΩ and $R_{xy} \approx -2.4$ kΩ. The resistance in both states is set by the DW configuration controlled by $I_{dc}$, which tunes the DW position and thus determines the resulting different $R_{xx}$ and $R_{xy}$ resistances (Methods). This represents one of the lowest magnetic domain switching currents reported to date, about six orders of magnitude smaller than the mA-scale currents typical of conventional spintronic devices [34,45], and comparable to or lower than those in moiré systems [35,36,46]. The corresponding switching current density, estimated as $J \approx 2I_{dc}/(wt_g) = 50$ A/cm$^2$ (assuming a DW length $w \approx 4$ µm in Fig. 5j and an R5G thickness $t_g$ =1.7 nm), is likewise orders of magnitude lower than in spintronic devices based on vdW magnetic materials and bulk ferromagnets [34,45,47], underscoring the exceptional efficiency of current-driven isospin domain manipulation in this system.

To elucidate the mechanism underlying the current-induced switching, we perform magnetic imaging in both the LR and HR states using $I_{ac}$ (see Methods and Extended Data Table 1 for measurement details). Figure 5d shows a $B_z^{ac}$ map of the LR state, revealing a blue domain above contact 4. This configuration can be understood as follows. After initialization at negative $B$ and sweeping to positive $B$, the majority phase is $K'{\downarrow}$ (blue in Fig. 5e schematic), with a hysteretic minority bubble nucleated near contact 4. Applying a negative current, $I_{dc} = -1.7$ nA, injects electrons at contact 4 and drives the expansion of this bubble. During the positive half-cycle of the *ac* modulation, $I = I_{dc} + |I_{ac}|$, the magnitude of the negative current is reduced, leading to a smaller bubble (red in Fig. 5e). In contrast, during the negative half-cycle, $I = I_{dc} - |I_{ac}|$, the increased injection further expands the bubble (Fig. 5f). The resulting $B_z^{ac}$ signal (Fig. 5d) reflects the difference between these two configurations, as illustrated schematically in Fig. 5g. Since contacts 2, 3, and 5 remain within the majority domain, the corresponding $R_{xx}$ and $R_{xy}$ stay low. The enhanced fluctuations in $R_{xx}$ and $R_{xy}$ at $I_{dc} = -1.7$ nA (Figs. 5b,c) highlight the sensitivity of the LR state to stochastic reconfigurations of the minority phase (Extended Data Fig. 5).

At $I_{dc} = +1.7$ nA (Fig. 5h), by contrast, the minority domain encompasses both contacts 3 and 4 (Figs. 5i,j). During the positive half-cycle of the *ac* modulation, $I = I_{dc} + |I_{ac}|$, the bubble shrinks due to electron drainage at contact 4 (red in Fig. 5i), while during the negative half-cycle, $I = I_{dc} - |I_{ac}|$, the bubble expands due to reduced drainage. The resulting $B_z^{ac}$ signal (Fig. 5h), reflecting the difference between the two configurations, is illustrated in Fig. 5k. With the highly resistive DW between contacts 2 and 3, contact 3 resides at the high potential of contact 4, while contact 2 is at the low potential of contact 1, yielding a large $V_{xx}$ and corresponding $R_{xx}$ across the intervening DW. Since contact 5 also lies in the majority domain at the potential of contact 1, the corresponding $R_{xy}$, measured between contacts 3 and 5, is comparable in magnitude to $R_{xx}$, but of opposite polarity (Figs. 5b,c).

A similar isospin domain configuration can also produce negative $R_{xx}$ (Extended Data Fig. 9), providing a natural explanation for previously enigmatic observations of NR in RMG [8,14]. In this regime, current-driven DW motion generates a negative differential response and hence an apparent negative longitudinal resistance (Methods).

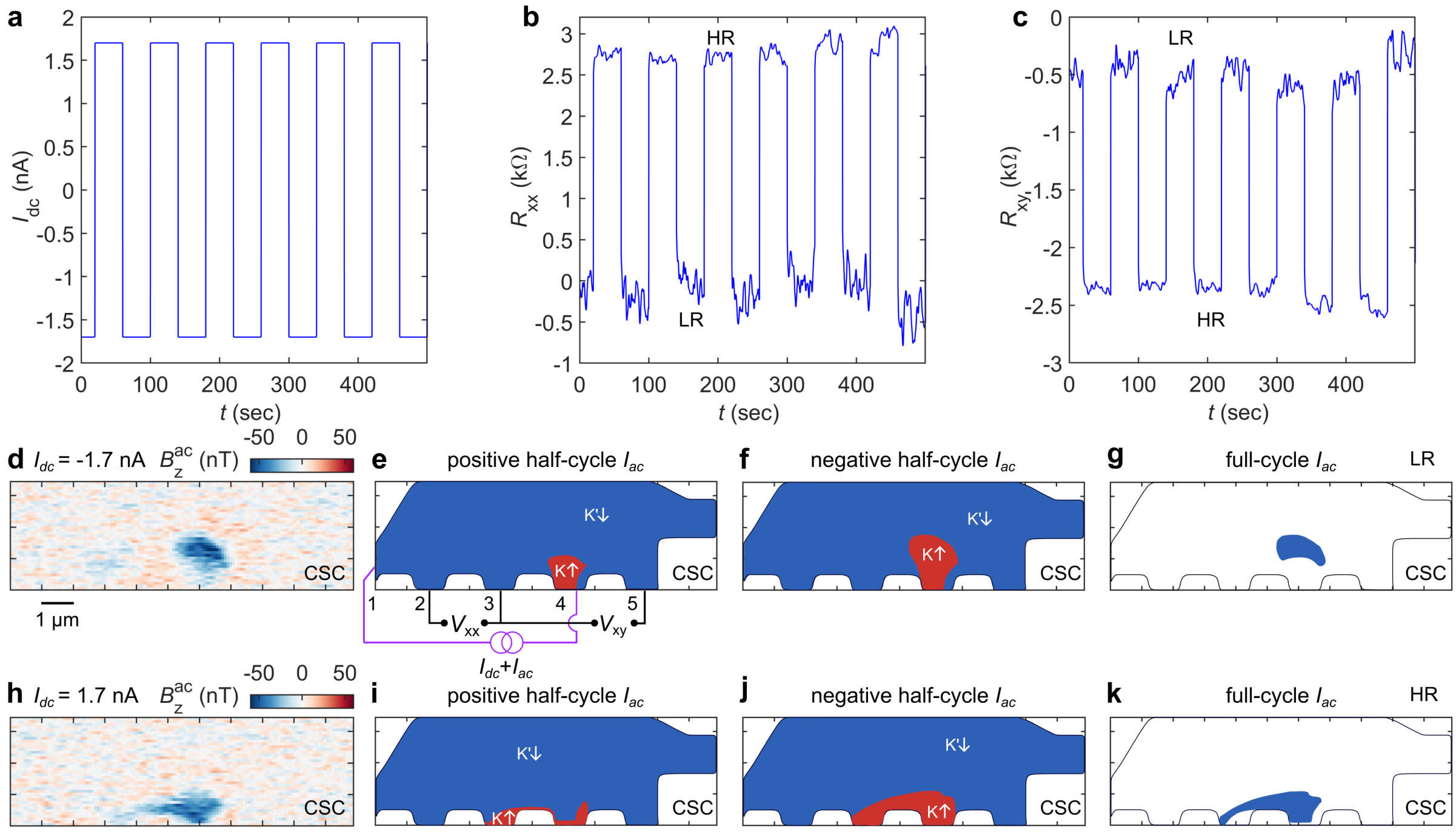

**Fig. 5. Current-driven switching of resistance and isospin domains in the CSC state. a,** Time sequence of *dc* current $I_{dc} = \pm 1.7$ nA applied between contacts 4 and 1 in device D1 in the CSC state ($T = 20$ mK, $D = -0.996$ V/nm, $n = 0.752 \times 10^{12}$ cm$^{-2}$, $B = 27$ mT following initialization at $B = -120$ mT). **b,** Corresponding $R_{xx}$ measured with $I_{ac} = 1$ nA, showing switching between the HR state ($R_{xx} \approx 2.7$ kΩ at $I_{dc} = +1.7$ nA) and the LR state ($R_{xx} \approx 0$ kΩ at $I_{dc} = -1.7$ nA). **c,** Same as (b), showing corresponding switching between $R_{xy} \approx -2.4$ kΩ at $I_{dc} = +1.7$ nA and $R_{xy} \approx 500$ Ω at $I_{dc} = -1.7$ nA. **d,** $B_z^{ac}$ map measured in the LR state at $I_{dc} = -1.7$ nA. **e,f,** Schematic isospin domain configurations during the positive (**e**) and negative (**f**) half-cycles of $I_{ac}$, where the minority domain shrinks and expands due to reduced and enhanced electron injection at contact 4, respectively. **g,** Difference map between (e) and (f), corresponding to the $B_z^{ac}$ signal in (d). Contacts 2, 3, and 5 lie within a single majority domain, resulting in LR response. **h,** Same as (d) but at $I_{dc} = +1.7$ nA, corresponding to the HR state. **i,j,** Schematic isospin domain configurations for the positive (**i**) and negative (**j**) half-cycle of $I_{ac}$. Contacts 2 and 5 remain in the majority domain, while contacts 3 and 4 form a minority domain, separated by a high-resistivity DW between contacts 2 and 3, resulting in HR configuration. **k,** Difference map between (i) and (j), reproducing the observed $B_z^{ac}$ signal in (h).

The observed current-driven DW dynamics are consistent with a mechanism dominated by linear momentum transfer from reflected electrons [33,48,49] (Methods). The large interfacial resistivity of the DW implies strongly suppressed transmission, placing the system in a non-adiabatic regime where electron reflection efficiently drives domain motion. Measurements in the presence of an in-plane field (Extended Data Fig. 10) further show that the DW resistance is reduced when the spin mismatch between neighboring domains is partially suppressed, indicating that both spin and valley contribute to the interfacial barrier. This establishes a regime of DW dynamics governed by coupled spin–valley degrees of freedom, distinct from conventional spin-transport [37,50] or valley-Hall [51,52] mechanisms. More broadly, these results demonstrate a highly efficient, reversible resistance-switching mechanism governed by current-controlled positioning of a single DW, pointing to isospintronics as a new paradigm for encoding and manipulating information through reconfigurable isospin textures in a chiral superconductor.

**Discussion**

The evolution of DW dynamics reveals a striking correlation between the onset density of spike-like hysteresis and the boundary of the CSC phase. Since the hysteresis transformation occurs both below and above $T_c$, it cannot be attributed to superconductivity itself, but instead points to a normal-state instability that emerges at $n_{DW}(D)$. This coincidence suggests that CSC develops out of a reconstructed parent state in which the energetics of isospin polarization are fundamentally modified (Methods and Extended Data Fig. 11), pointing to an underlying symmetry-breaking transition that reshapes the energy landscape from which CSC emerges and calls for further theoretical investigation.

While the proliferation of DWs is largely inherited from the parent 1/4M phase, several important differences emerge upon entering the CSC phase. Below $T_c$, the HR state becomes more stable and persists over a wider magnetic-field range (Figs. 2i-l), while larger currents are required to drive DW motion (Extended Data Fig. 8). This behavior can be understood if the SC order parameter increases the DW line tension and pinning by introducing an additional energy cost associated with suppressing the SC condensate within the DW (Methods). Moreover, this additional pinning provides a natural explanation for the subtle hysteresis observed upon sweeping $B$ in the LR state in CSC phase (Figs. 2i,j and Extended Data Figs. 3 and 9). In this picture, in addition to separating areas of opposite chirality, DW may be regarded as a region where the superconducting order is locally suppressed. This interpretation is supported by the observation that, in the HR state, DWs remain highly resistive even at very low bias currents, with no detectable Josephson critical current, indicating the absence of coherent tunneling between adjacent domains. Such behavior is consistent with superconducting order parameters that differ across the wall in phase or chirality, suggesting a nontrivial internal structure of the superconducting state. However, while this work establishes macroscopic TRS breaking and CSC in pentalayer graphene, determining whether the superconducting order parameter itself is intrinsically chiral requires further investigation.

Importantly, the coexistence of superconductivity with oppositely polarized chiral domains has no analogue in any other superconductors. By contrast, CSC in RMG combines superconductivity with a bistable, magnetically ordered state, a defining hallmark of this phase. These findings establish CSC in RMG as a reconfigurable state emerging from the interplay between superconductivity and an underlying isospin-polarized order, in which the chiral domain structure is inherited from the parent phase. More broadly, the ability to directly image and manipulate isospin domains with ultra-low currents opens a route to coupling superconductivity to real-space magnetic textures, establishing rhombohedral graphene as a model platform for isospintronics, where electronic functionality can be controlled through the interplay of spin–valley degrees of freedom and superconducting pairing.

**Acknowledgements**

We thank Ido Ben Amram for improving the neural network model and reconstructing the differential magnetization. This work was co-funded by the Minerva Stiftung with funding from the Federal German Ministry for Education and Research, by the United States - Israel Binational Science Foundation (BSF) grant No 2022013, by the Knell Family Institute for Artificial Intelligence, and by the European Union (ERC, MoireMultiProbe - 101089714). Views and opinions expressed are however those of the author(s) only and do not necessarily reflect those of the European Union or the European Research Council. Neither the European Union nor the granting authority can be held responsible for them. E.Z. acknowledges the support of the Goldfield Family Charitable Trust and the Tom and Mary Beck Center for Advanced and Intelligent Materials at the Weizmann Institute of Science. G.S. acknowledges support from the Walter Burke Institute for Theoretical Physics at Caltech, and from the Yad Hanadiv Foundation through the Rothschild fellowship. K.W. and T.T. acknowledge support from the JSPS KAKENHI (Grant Numbers 21H05233 and 23H02052), the CREST (JPMJCR24A5), JST and World Premier International Research Center Initiative (WPI), MEXT, Japan.

**Author contributions**

S.D. performed the local magnetization and transport measurements and analyzed the data. N.A. designed and built the scanning SOT microscope. T.H. fabricated the samples and characterized them with electron transport measurements. Y.Z. performed COMSOL numerical simulations. G.S. developed the theory of domain wall energetics. N.S.K, S.D., and Y.M. fabricated the SOT and the tuning fork, and M.E.H. developed the SOT readout. K.W. and T.T. provided the hBN crystals. E.Z., S.D. and N.A. designed the experiment. E.Z., S.D., and L.J. wrote the original manuscript. All authors participated in discussions and revisions of the manuscript.

**Competing interests:** The authors declare no competing interests.

**Data availability:** The data that supports the findings of this study are available from the corresponding authors on reasonable request.

**Code availability:** The calculation codes used in this study are available from the corresponding authors on request.

## Methods

### Device fabrication

The graphene and hBN flakes were prepared by mechanical exfoliation onto $SiO_2$/Si substrates. The rhombohedral domains of pentalayer graphene were identified and confirmed using IR camera, near-field infrared microscopy, and Raman spectroscopy and isolated by cutting with a femtosecond laser. The van der Waals heterostructure was made following a dry transfer procedure. We picked up the top hBN, graphite, middle hBN and graphene using poly (bisphenol A carbonate) film on polydimethylsiloxane (PDMS) and landed it on a prepared bottom stack consisting of an hBN and graphite bottom gate. The device was then etched into a multi-terminal structure using e-beam lithography and reactive-ion etching. We deposited Cr–Au for electrical connections to the source, drain and gate electrodes.

### Transport measurements

Four-probe transport measurements were performed using standard lock-in techniques down to the base temperature of dilution refrigerator ($T = 20$ mK). As both devices had five contacts on a single side (Fig. 1a,b), in order to allow simultaneous measurement of both $R_{xx}$ and $R_{xy}$, *ac* current was sourced between contacts 4 (high) and 1 (low) and the *ac* voltages $V_{xx} = V_3 - V_2$ and $V_{xy} = V_5 - V_3$ were measured across contacts 3-2 and 5-3, respectively. Contacts 1 and 4 exhibited the lowest contact resistance, and were therefore used in most measurements, typically using $I_{ac} = 1$ nA rms and $f$ =11.333 Hz, unless stated otherwise. The carrier density $n = (C_{tg}V_{tg} + C_{bg}V_{bg})/e$ and displacement field $D \equiv (C_{tg}V_{tg} - C_{bg}V_{bg})/2\epsilon_0$ were independently controlled via the top and bottom gate voltages $V_{tg}$ and $V_{bg}$. Here, $C_{tg}$ and $C_{bg}$ are the geometric capacitances per unit area of top and bottom gates, $e$ is the elementary charge, and $\epsilon_0$ is vacuum permittivity. Extended Data Figs. 1,2 show additional transport data of devices D1 and D2.

### SOT measurements

Two indium-based SQUID-on-tip (SOT) sensors were fabricated on the apex of shape quartz pipettes using thermal evaporation, following previously reported procedures [53–55]. The SOTs have diameters of 220 nm, as determined from their interference patterns, and operate in magnetic fields up to 0.45 T with a field sensitivity of 10 nT/$\sqrt{\mathrm{Hz}}$. The tip-sample distance (Extended Data Table 1) was controlled by attaching the SOT to a quartz tuning fork [56] vibrating at its resonance frequency $\approx$ 32 kHz. The SOT signal was read out using a cryogenic SQUID series array amplifier (SSAA) [57]. Local magnetic imaging was performed in a dilution refrigerator with base temperature $T = 20$ mK. To enhance the signal-to noise ratio, all SOT measurements were carried out in *ac* mode, using either $V_{tg}^{ac}$, $V_{bg}^{ac}$, or $I_{ac}$ modulation in the frequency range of 400 Hz to 2.1 kHz. In measurements involving current-driven SOT imaging, the high-frequency $I_{ac}$ simultaneously drives DW dynamics and enables concurrent measurements of $R_{xx}$ and $R_{xy}$. Low-pass filters with a cutoff frequency of $\sim$1 kHz were installed on all electrical lines to the divides. As a result, the $R_{xx}$ and $R_{xy}$ signals measured at high frequency during SOT imaging should be regarded as qualitative. Measurement parameters for the different figures are summarized in Extended Data Table 1.

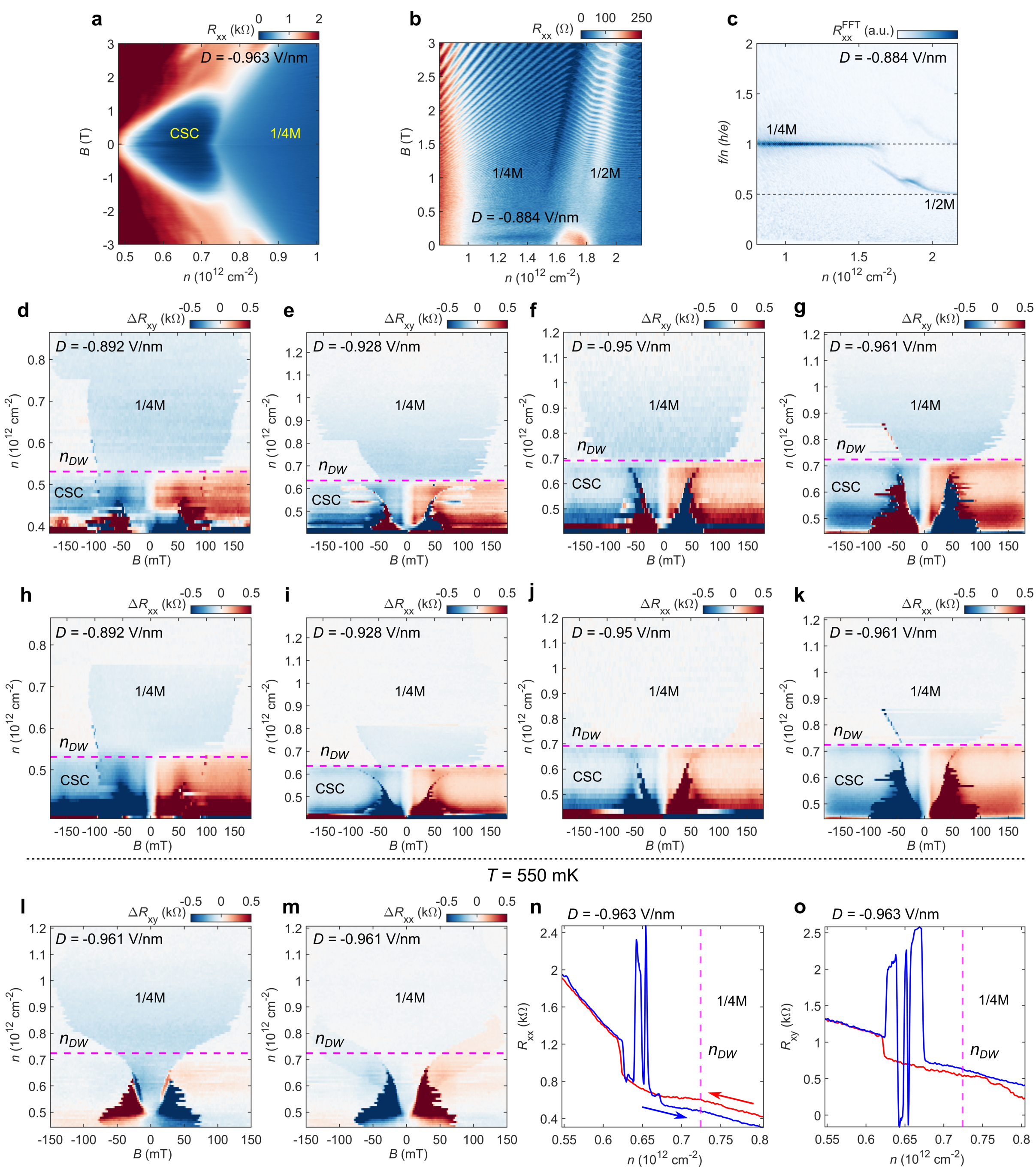

**Extended Data Fig. 1. Transport properties of R5G device D1. a,** $R_{xx}(n, B)$ measured at $D = -0.963$ V/nm and $T = 20$ mK. **b,** Landau fan diagram of $R_{xx}(n, B)$ at $D = -0.884$ V/nm and $T = 20$ mK. **c,** Corresponding fast Fourier transform (FFT) of the quantum oscillations computed from $R_{xx}(1/B)$ in (c) as a function of $n$, indicating 1/4M to 1/2M transition. **d-g,** Magnetic field hysteresis $\Delta R_{xy}(B, n) = R_{xy}(B\uparrow, n) - R_{xy}(B\downarrow, n)$ at $T = 20$ mK measured at $D = -0.892$ (d), $-0.928$ (e), $-0.95$ (f), and $-0.961$ (g) V/nm. Magenta dashed lines mark $n_{DW}$, the onset density of spike-like hysteresis due to proliferation of DWs. The corresponding $n_{DW}(D)$ values are marked by magenta squares in Figs. 1d,e, coinciding with the high-density boundary of the CSC region. **h-k,** Corresponding magnetic field hysteresis of $\Delta R_{xx}(B, n) = R_{xx}(B\uparrow, n) - R_{xx}(B\downarrow, n)$. **l,m,** Same as (g,k) but measured at $T = 550$ mK. **n,o,** Hysteresis of $R_{xx}$ (n) and $R_{xy}$ (o) as a function of $n$, measured at $B = 0$ mT, $D = -0.963$ V/nm, and $T = 550$ mK. The vertical magenta dashed lines mark $n_{DW}(D)$.

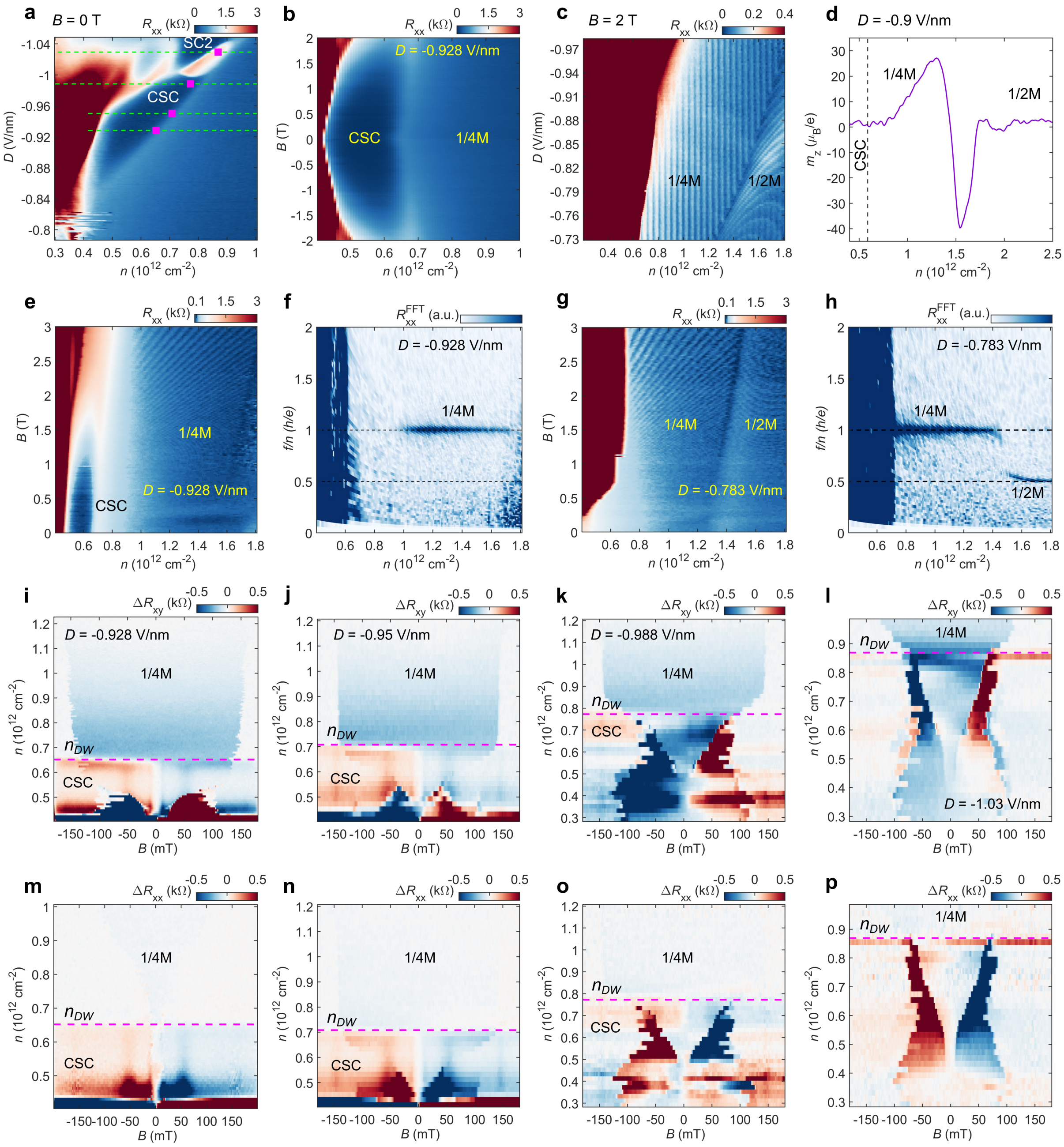

**Extended Data Fig. 2. Transport properties of R5G device D2. a,** $R_{xx}(n, D)$ measured at $B = 0$ T and $T = 20$ mK, showing two SC pockets CSC and SC2. The magenta marks indicate $n_{DW}(D)$. **b,** $R_{xx}(n, B)$ measured at $D = -0.928$ V/nm and $T = 20$ mK. **c,** $R_{xx}(n, D)$ measured at $B = 2$ T, showing vertical quantum oscillation lines with degeneracy corresponding to 1/4M with a single Fermi surface and tilted quantum oscillation lines in 1/2M phase. **d,** Local differential magnetization $m_z = dM_z/dn$, computed from $B_z^{ac}$ measured at a single location in device D2 using bottom-gate modulation $V_{bg}^{ac} = 200$ mV at $B = 58$ mT and $T = 20$ mK. In the 1/4M phase, $m_z$ strongly decreases with reducing $n$. **e,** Landau fan diagram $R_{xx}(n, B)$ measured at $D = -0.928$ V/nm showing quantum oscillations at high fields. **f,** Corresponding FFT of the quantum oscillations computed from $R_{xx}(1/B)$ in (e) showing oscillation frequency corresponding to 1/4M phase. **g,** Landau fan diagram $R_{xx}(n, B)$ measured at $D = -0.783$ V/nm below CSC region. **h,** Corresponding FFT showing 1/4M to 1/2M transition. **i-l,** Magnetic field hysteresis $\Delta R_{xy}(n, B) = R_{xy}(B\uparrow, n) - R_{xy}(B\downarrow, n)$ measured at $T = 20$ mK at $D = -0.928$ (i), $-0.95$ (j), $-0.988$ (k), and $-1.03$ V/nm (l) along the green dashed lines in (a). Magenta dashed lines mark $n_{DW}$ below which the hysteresis patterns change qualitatively. The corresponding $n_{DW}(D)$ values are marked by magenta squares in (a). **m-p,** Corresponding magnetic field hysteresis of $\Delta R_{xx}(B,n) = R_{xx}(B\uparrow,n) - R_{xx}(B\downarrow,n)$.

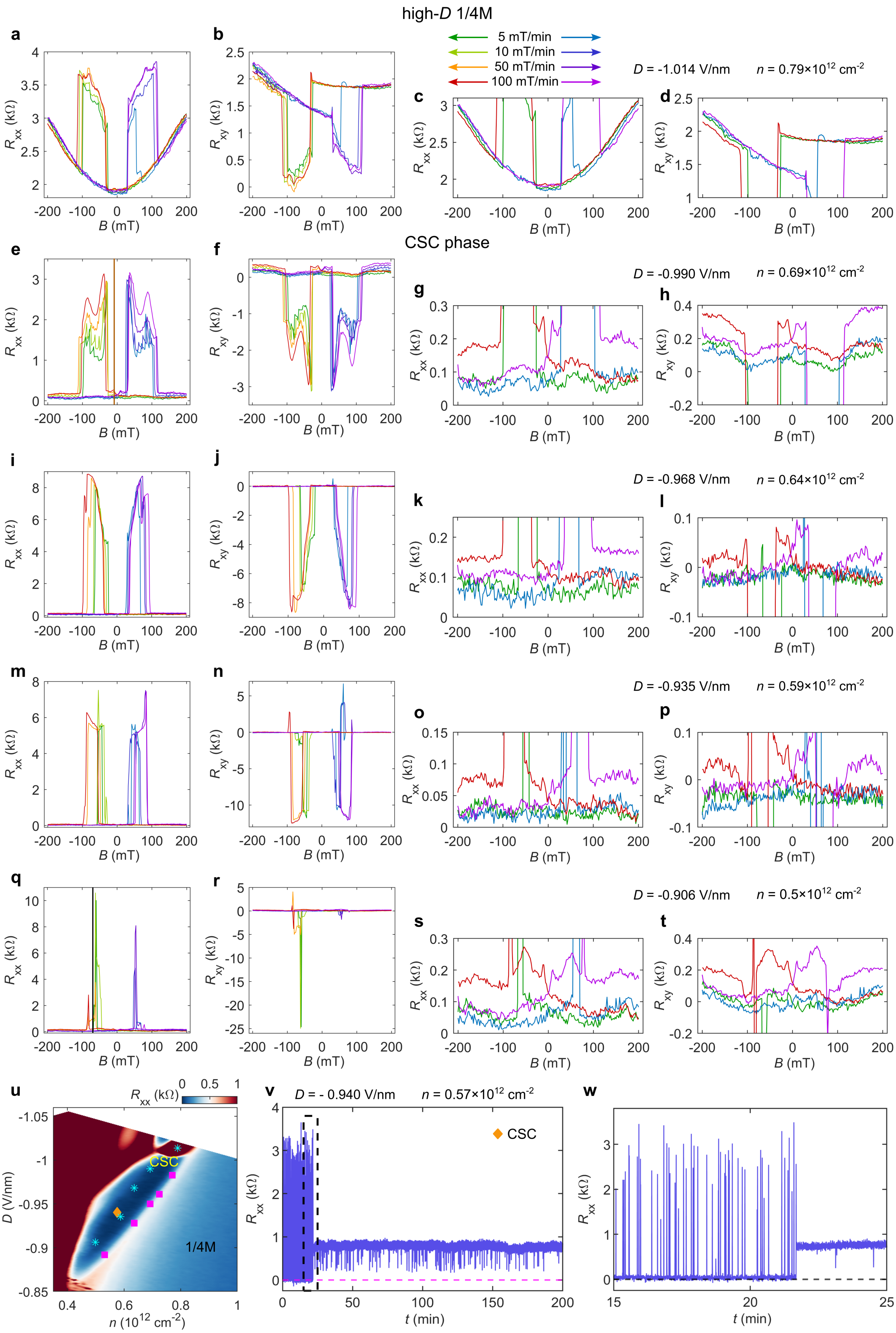

**Extended Data Fig. 3. Magnetic-field sweep-rate and phase-diagram dependence of transport hysteresis. a,b,** Magnetic-field hysteresis of $R_{xx}$ (**a**) and $R_{xy}$ (**b**) measured in high-$D$ 1/4M phase ($D = -1.014$ V/nm and $n = 0.79\times10^{12}$ cm$^{-2}$; top cyan star in (u)) in device D1 at $T = 20$ mK upon sweeping $B$ from 200 to $-200$ mT and back at four different ramp rates between 5 and 100 mT/min. **c,d,** Zoomed-in view of the LR-state hysteresis in $R_{xx}$ (**c**) and $R_{xy}$ (**d**) for the minimum (5 mT/min) and maximum (100 mT/min) ramp rates. **e-t,** Same measurements as in (a-d), performed across the phase diagram in the CSC phase at progressively decreasing $|D|$ values (cyan stars in (u)). **u,** Zoomed-in $R_{xx}(n, D)$ phase diagram reproduced from Fig. 1d, indicating the measurement points (cyan stars). Magenta markers denote the DW proliferation onset density $n_{DW}(D)$. **v,** Temporal evolution of $R_{xx}(t)$ measured continuously over more than 3 hours in the CSC phase ($D = -0.940$ V/nm and $n = 0.57\times10^{12}$ cm$^{-2}$, orange diamond in (u)) at $B = 11$ mT and $T = 20$ mK, showing large stochastic fluctuations in an initially ZR state, followed by a random transition to a fluctuating LR state. **w,** Zoom-in of $R_{xx}(t)$ for $15 < t < 25$ min (dashed box in (v)), showing a spontaneous transition from ZR to LR state.

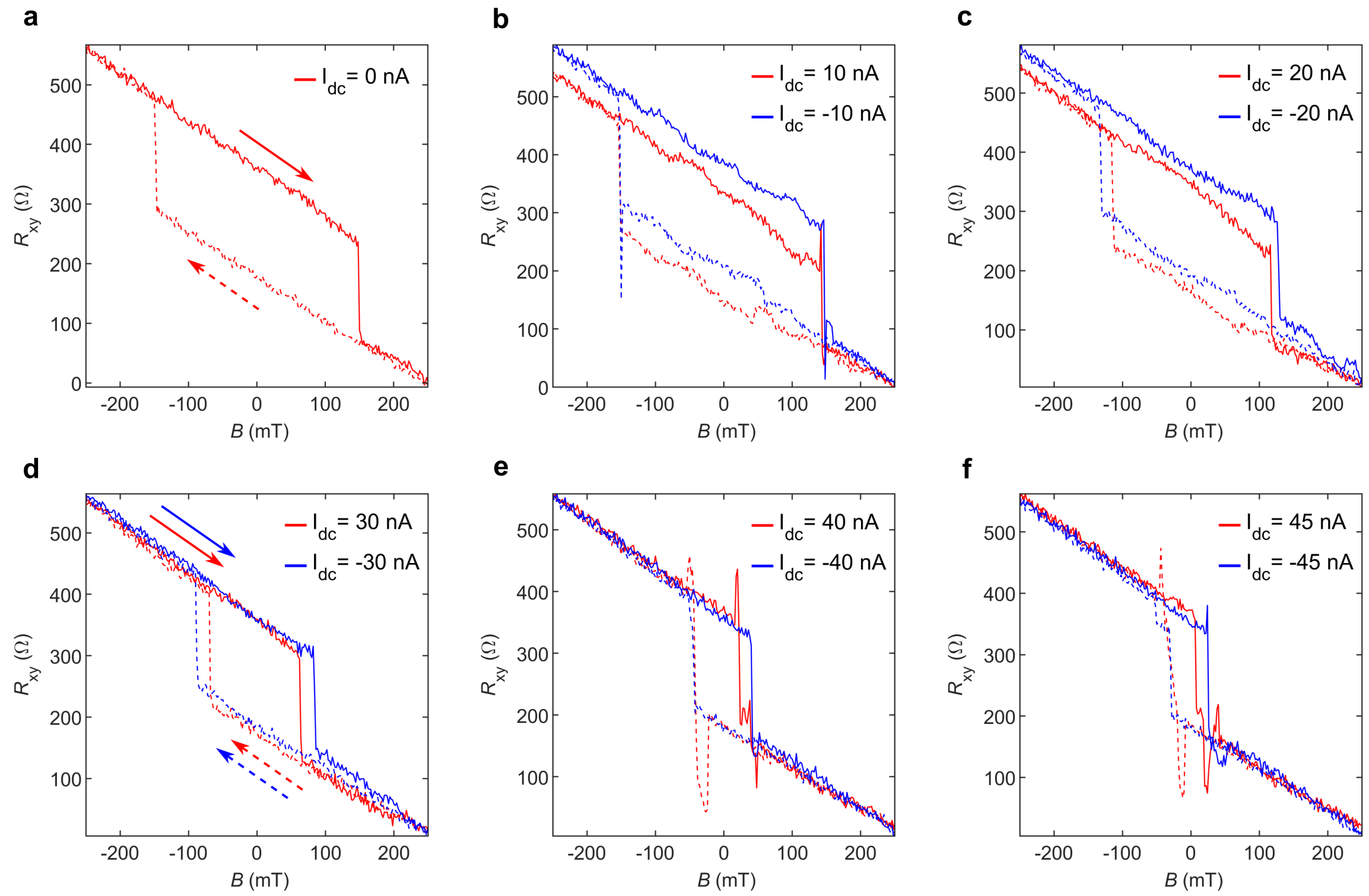

**Extended Data Fig. 4. Current dependence of magnetic hysteresis width in the 1/4M phase. a-f,** Hysteresis of $R_{xy}$ measured upon increasing (solid) and decreasing (dashed) magnetic field in the presence of applied *dc* current $I_{dc}$ in device D2 at $D = -0.95$ V/nm, $n = 0.8\times10^{12}$ cm$^{-2}$, and $T = 20$ mK in the 1/4M phase. As $I_{dc}$ increases from 0 (**a**) to 45 nA (**f**), the coercive field $B_{co}$ is significantly suppressed. Positive (red) and negative (blue) *dc* currents produce a similar reduction in $B_{co}$.

### Current-driven isospin domain dynamics

Extended Data Fig. 5 shows that the DW dynamics are accompanied by significant stochastic domain reconfigurations, leading to large resistance fluctuations and noise (Extended Data Figs. 3v,w). In addition, the stochastic nature of the DW motion strongly affects the intensity of $B_z^{ac}$. This is evident, for example, in Extended Data Figs. 8e,f, where the faint blue bubble near contact 1 contrasts with the much more intense red bubble near contact 4 (see also Extended Data Fig. 8h). This difference reflects the distinct character of the DW motion: near contact 1, the DW undergoes large but highly stochastic excursions, whereas near contact 4 the motion is more confined and reproducible from cycle to cycle. Since each pixel in the $B_z^{ac}$ map represents an average over approximately 300 *ac* cycles, such stochastic variability reduces the magnitude of the time-averaged signal. As the current amplitude is increased, the DW motion becomes progressively more deterministic. In this regime, the contrast between different bubbles diminishes, and a saturated time-averaged magnetization reversal of $\Delta M_z = 2M_z$ is achieved within the regions swept by the DW, as reflected in the enhancement and eventual saturation of the bubble intensities in Extended Data Figs. 8g–i.

Notably, the spatial extent of the DW excursions saturates at high current (compare Extended Data Figs. 8h and 8i). This behavior reflects the device geometry, in which both source and drain contacts reside on the same edge. The expansion of the minority bubble is driven by carrier injection at the source contact; once the bubble extends to encompass both source and drain, further growth is limited by carrier drainage at the opposite contact. This saturation of the excursion range at higher current also promotes more deterministic DW oscillations, thereby maximizing the $B_z^{ac}$ signal.

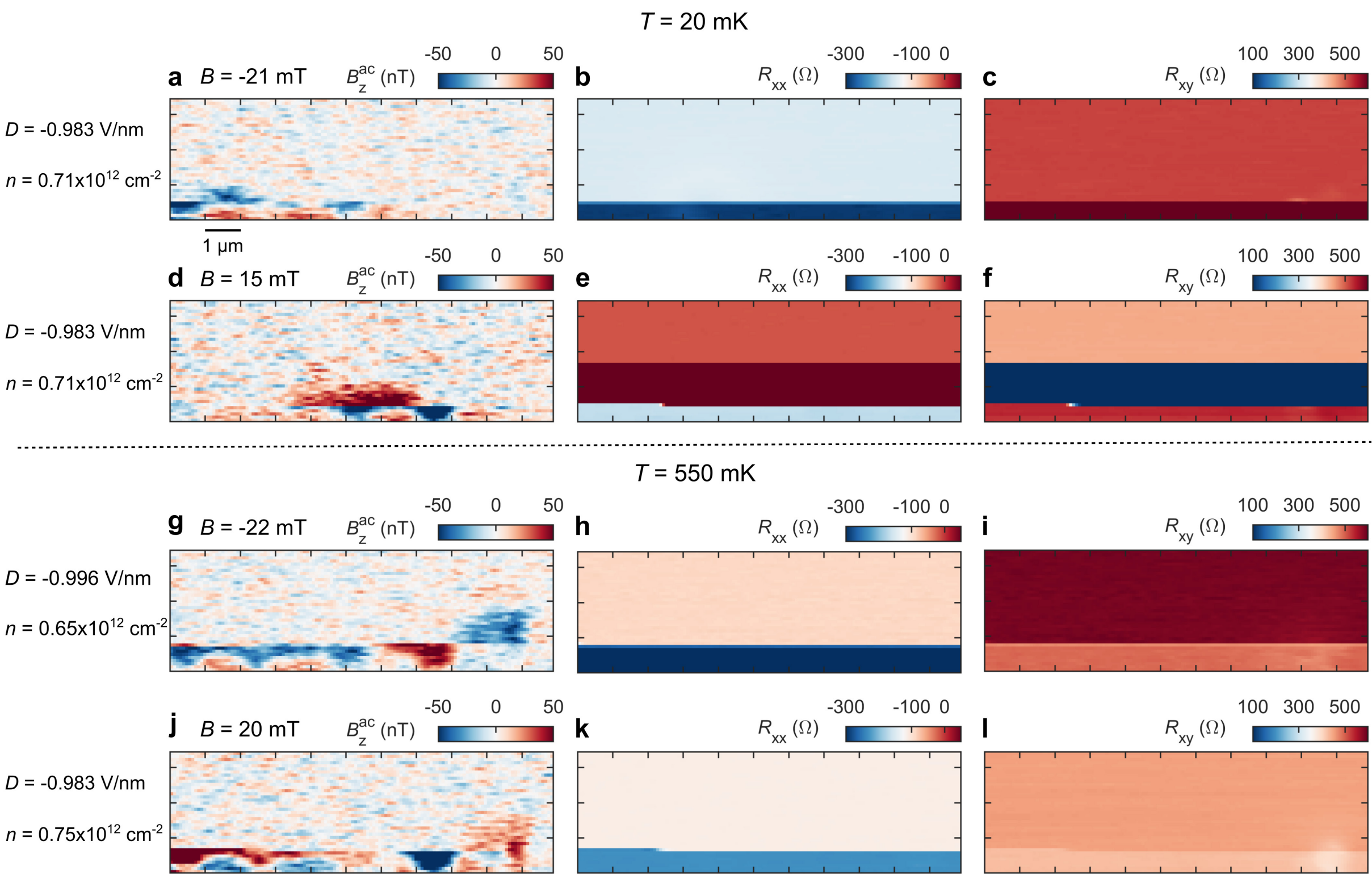

**Extended Data Fig. 5. Stochastic reconfigurations of isospin domains in the HR state in the CSC region at $T = 20$ mK and 550 mK. a,** $B_z^{ac}$ measured in device D1 using $I_{ac} = 10$ nA at $f = 1.181$ kHz, $D = -0.983$ V/nm, $n = 0.71\times10^{12}$ cm$^{-2}$, $B = -21$ mT, and $T = 20$ mK in the HR state in the CSC phase. Imaging is performed by SOT scanning along the fast $x$ axis and slow $y$ axis, starting from the lower-left corner. The isospin domain configuration changes abruptly after several scan lines. **b,c,** Corresponding $R_{xx}$ (b) and $R_{xy}$ (c) measured simultaneously during scanning using the same high-frequency $I_{ac}$. Each pixel in the resistance images reflects the instantaneous values of $R_{xx}(t)$ and $R_{xy}(t)$ at the time $t$ corresponding to the SOT position in (a). Sharp jumps in $R_{xx}$ and $R_{xy}$ reflect stochastic reconfigurations of the isospin domains observed in (a). The values of $R_{xx}$ and $R_{xy}$ are strongly influenced by low-pass filters on the electrical lines, which suppress the amplitudes and change the phase of the $V_{xx}$ and $V_{xy}$ voltages measured at high frequency, and should therefore be regarded only as qualitative indicators of domain dynamics. **d-f,** Same as (a-c) at $n = 0.75\times10^{12}$ cm$^{-2}$ and $B = 15$ mT. **g-i,** Same as (a-c) at $D = -0.996$ V/nm, $n = 0.65\times10^{12}$ cm$^{-2}$, $B = -22$ mT, and $T = 550$ mK $> T_c$ using $I_{ac} = 11.5$ nA. **j-l,** Same as (g-i) at $D = -0.983$ V/nm, $n = 0.75\times10^{12}$ cm$^{-2}$, and $B = 20$ mT.

In conventional spin-torque spintronic devices and vdW ferromagnets, a constant current drives DW motion at a steady velocity [37,49,58]. In contrast, we find that electron injection produces a static displacement of the DW, leading to an expansion of the minority domains, as demonstrated in Fig. 4 and Extended Data Fig. 6 using *dc* current. To estimate the timescale required to reach this static configuration, we performed $B_z^{ac}$ imaging using *ac* currents over a frequency range of $f$ = 633 to 1,533 Hz, within the experimentally accessible range. If DW propagation were slow compared to the *ac* period, the extent of the DW excursions would be expected to decrease approximately as $1/f$. Conversely, if the static configuration is reached quasi-instantaneously, the observed domain size should be independent of frequency.

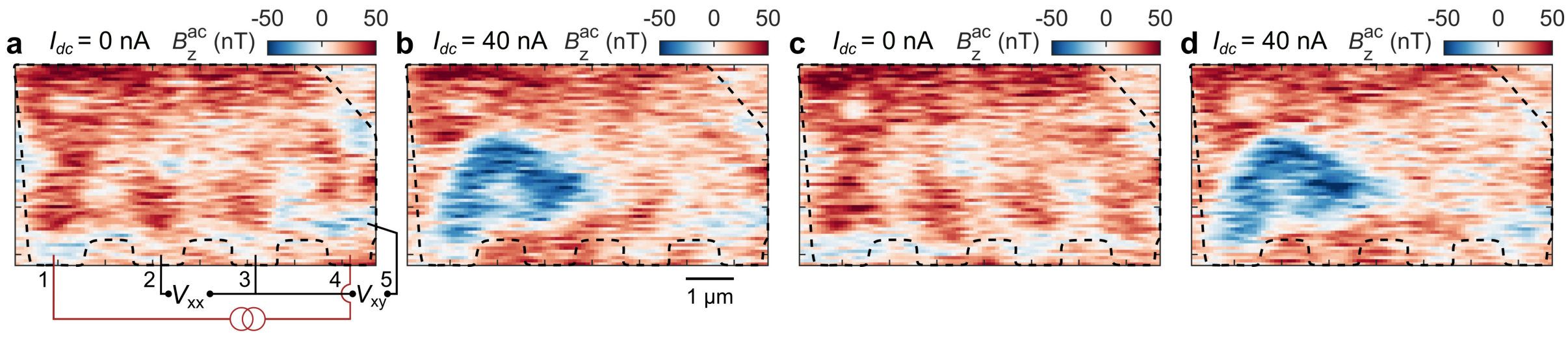

**Extended Data Fig. 6. Current-driven reversible switching of isospin domains. a,** $B_z^{ac}$ map in device D2 measured using $V_{bg}^{ac}$ modulation at $D = -1.03$ V/nm, $n$ = 0.422×10$^{12}$ cm$^{-2}$, $B$ = 13 mT, $T$ = 20 mK, and $I_{dc}$ = 0 nA in the high-$D$ 1/4M phase, reproduced from Fig. 4e. **b,** Same as (a) after increasing the current to $I_{dc}$ = 40 nA. A minority bubble is formed at contact 1, where electrons are injected. **c,** Same as (b) after reducing the current back to $I_{dc}$ = 0 nA. The bubble shrinks reversibly. **d,** Same as (c) after increasing the current again to $I_{dc}$ = 40 nA. The bubble re-expands, reproducing the configuration in (b).

The results in Extended Data Fig. 7 support the latter scenario, indicating that the steady-state domain configuration is established on timescales shorter than ~0.3 ms. Notably, while the spatial extent of the DW excursions shows little dependence on frequency, the amplitude of $B_z^{ac}$ decreases with increasing $f$. This reduction reflects the stochastic nature of DW motion in the presence of disorder and fluctuations. Although the maximal DW excursion can, in principle, be reached rapidly, the probability of reproducibly attaining it within each *ac* cycle decreases as the cycle period becomes shorter.

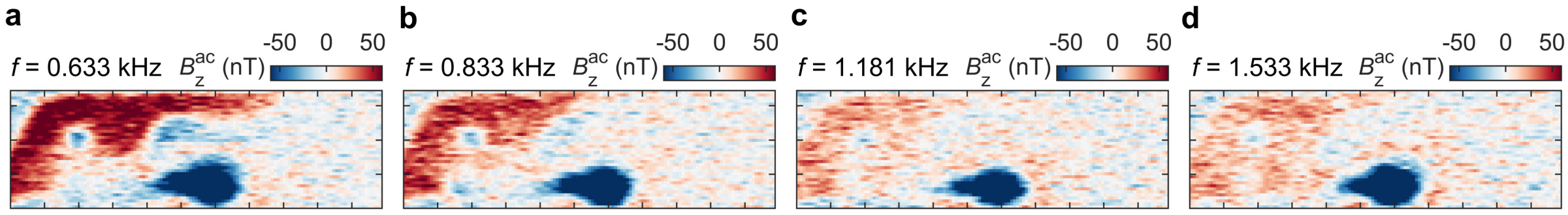

**Extended Data Fig. 7. Frequency dependence of DW dynamics. a-d,** $B_z^{ac}$ maps in the HR state in the CSC phase of device D1 at $D = -0.996$ V/nm, $n$ = 0.75×10$^{12}$ cm$^{-2}$, $B$ = 26 mT, and $T$ = 20 mK, measured using $I_{ac}$ = 6.1 nA at different frequencies: $f$ = 633 Hz (a), 833 Hz (b), 1.181 kHz (c), and 1.533 kHz (d). The spatial extent of the bipolar minority domains is largely unaffected by frequency, whereas the signal amplitude decreases with increasing $f$, indicating an enhanced role of stochastic dynamics at shorter timescales.

**DW dynamics governed by linear momentum transfer**

Current-driven DW motion can arise from two distinct microscopic mechanisms. In the conventional spin-transfer-torque mechanism, electrons traverse the domain wall while their spin adiabatically follows the local order parameter, thereby transferring angular momentum to the wall. This mechanism dominates in metallic ferromagnets with wide and highly transparent DWs [37]. In contrast, when electron transmission across the wall is suppressed, it has been proposed that the DW can be moved by the linear momentum transfer from reflected electrons [33,48,49].

Our observations indicate that rhombohedral graphene provides the first realization of DW motion dominated by linear momentum transfer. The measured interfacial resistivity of the isospin DW, $\rho_{DW} \approx$ 41 kΩ·µm, implies strongly suppressed electron transmission across the interface. Moreover, since the two domains correspond to opposite valleys of the Brillouin zone, the isospin order parameter is expected to change on a length scale comparable to the lattice constant to mix the $K$ and $K'$ states, resulting in a DW width of only a few lattice constants [59]. Such narrow walls place the system deep in the non-adiabatic limit, where electrons cannot follow the rapid spatial variation of the order parameter and are therefore strongly reflected. Transmission is further suppressed by the requirement of simultaneous intervalley scattering and spin flip.

To probe the role of spin in this process, we measured the hysteretic transport in the presence of an in-plane magnetic field $B_\parallel$ (Extended Data Fig. 10). While the overall hysteretic behavior and spike-like HR states remain qualitatively unchanged, the amplitude of the resistance spikes in both $R_{xx}$ and $|R_{xy}|$ decreases rapidly with increasing $B_\parallel$, saturating at a reduced value for $|B_\parallel| \gtrsim 0.1$ T. Since $B_\parallel$ tends to align the spins in neighboring domains with opposite isospin polarization, this reduction indicates that the large $\rho_{\mathrm{DW}}$ at $B_\parallel = 0$ T arises from the combined mismatch of both valley and spin degrees of freedom, whereas at high $B_\parallel$, $\rho_{\mathrm{DW}}$ is primarily governed by valley momentum mismatch. This highlights the inherently multicomponent nature of isospin transport, positioning isospintronics as a framework for electronic functionality based on the coupled control of multiple quantum degrees of freedom, extending concepts developed in spintronics [37,45,47,58,60,61] and valleytronics [51,52,62–64] in vdW materials.

To quantitatively estimate the electron transmission probability across the DW, we employ a Landauer approach. For a 2D conductor, $\rho_{\mathrm{DW}}$ is related to the transmission probability $\beta$ as

$$\rho_{\mathrm{DW}} = \frac{1}{\beta}\frac{h}{e^2}\frac{\pi}{k_F},$$

where $h$ is Planck constant and $e$ is the elementary charge. In the 1/4M phase, the Fermi momentum is $k_F = \sqrt{4\pi n} \approx$ 2.5×10$^6$ cm$^{-1}$ for $n \approx$ 0.5×10$^{12}$ cm$^{-2}$ (Fig. 2). Using $\rho_{DW} \approx$ 40 kΩ·µm, we obtain $\beta \approx$ 8×10$^{-3}$, indicating that the DW is nearly perfectly reflecting. In this limit, the current-induced force $f_{\mathrm{DW}}$ is governed by momentum transfer: each normally reflected electron transfers momentum $\Delta p = 2\hbar k_F$ to the DW, yielding

$$f_{\mathrm{DW}} \approx \frac{2\hbar k_F}{e} J_\perp,$$

where $J_\perp$ is current density normal to the DW. For $I_{dc} = 30$ nA (Fig. 4c), the minority bubble has a circumference of ≈7 µm, corresponding to $J_\perp \approx$ 4 nA/µm and a force $f_{\mathrm{DW}} \approx$ 1.3×10$^{-15}$ N/µm. Since the bubble remains stationary, this force must be balanced by pinning, providing an estimate of the DW pinning force, $f_{\mathrm{pin}} \approx f_{\mathrm{DW}}$.

An independent estimate of $f_{\mathrm{pin}}$ can be obtained from the coercive field $B_{co}$ by equating the force on the DW to the gradient of the Zeeman energy difference between oppositely polarized domains, $f_{\mathrm{pin}} = \frac{dU}{dx} = 2M_z B_{co}$. Using $B_{co} \approx$ 50 mT and a magnetization of ≈3 µ$_B$ per electron at $n \approx$ 0.4×10$^{12}$ cm$^{-2}$ (Fig. 4g) yields $f_{\mathrm{pin}} \approx$ 1×10$^{-14}$ N/µm. The order-of-magnitude agreement between these two independent estimates provides support for the momentum-transfer mechanism governing the DW dynamics.

Interestingly, the qualitative DW dynamics remain largely unchanged upon cooling the device into the CSC phase. The $B_z^{ac}$ patterns measured at $T = 20$ mK (Extended Data Figs. 8a-f) are qualitatively similar to those at $T = 550$ mK $> T_c$ (Extended Data Figs. 8g-i), exhibiting bipolar DW excursions driven by the $ac$ current. This demonstrates that the coexistence of oppositely polarized domains is primarily governed by the isospin order parameter of the parent 1/4M phase. Notably, however, substantially larger $I_{ac}$ is required to achieve comparable DW excursions in the CSC state. For example, the spatial extent and magnitude of the $B_z^{ac}$ patterns in Extended Data Figs. 8f and 8h are similar, yet the required current at 20 mK ($I_{ac} = 6.55$ nA) is approximately four times larger than that at 550 mK (1.63 nA). This increase indicates stronger DW pinning in the CSC phase. This interpretation is consistent with the larger coercive fields and the wider stability range of the HR states observed in the CSC regime (Figs. 2i,j) compared to the data at $T = 1$ K (Figs. 2k,l).

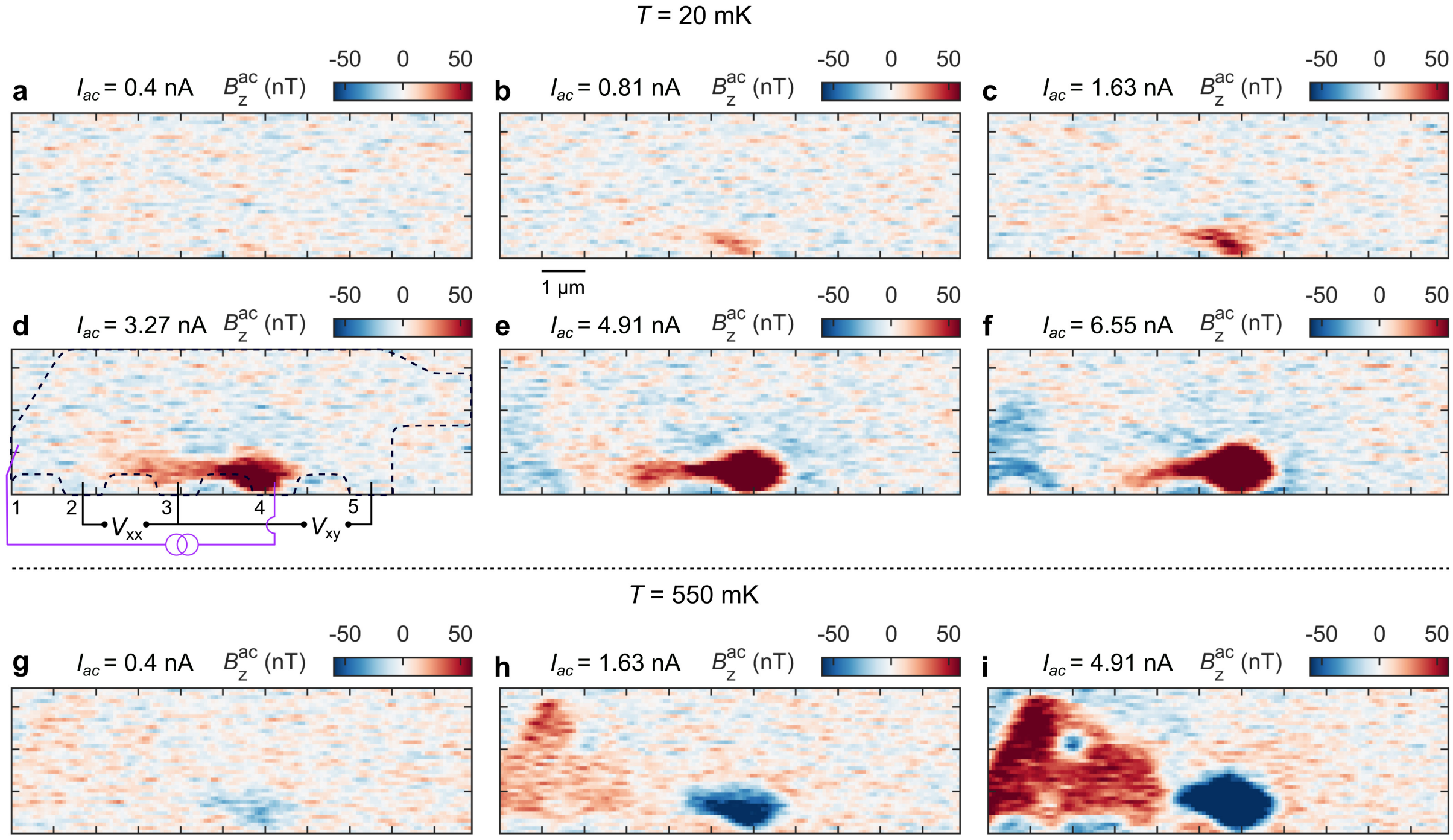

**Extended Data Fig. 8. Dependence of isospin domain dynamics on *ac* current amplitude in the CSC region at $T = 20$ and 550 mK. a-f,** $B_z^{ac}$ maps measured as function of increasing $I_{ac}$ amplitude in the HR state in the CSC phase of device D1 at $D = -0.996$ V/nm, $n = 0.75 \times 10^{12}$ cm$^{-2}$, $B = -20$ mT, and $T = 20$ mK. **a,** $B_z^{ac}$ at $I_{ac} = 0.4$ nA, showing the onset of DW motion near current-injection contact 4, appearing as a barely resolvable small bubble (red). **b,** $I_{ac} = 0.81$ nA, where DW dynamics near contact 4 become clearly resolved (red), corresponding to expansion and contraction of a minority isospin bubble. **c,d,** Same as (b) at $I_{ac} = 1.63$ nA and 3.27 nA, respectively; the DW motion progressively extends to encompass contacts 3 and 4. **e,** $I_{ac} = 4.91$ nA, where an additional minority bubble nucleates near contact 1 (blue), oscillating out of phase with the bubble near contacts 3 and 4 (red). **f,** $I_{ac} = 6.55$ nA, showing pronounced bipolar, out-of-phase expansion and contraction of minority bubbles near contacts 1 and 3–4. **g-i,** Same as (a-f), but measured at $B = 19.5$ mT and $T = 550$ mK. The minority bubbles appear with opposite contrast due to the reversed magnetic field polarity. The spatial extent of the response is significantly enhanced at 550 mK compared to 20 mK for similar $I_{ac}$, indicating weaker DW pinning in the normal state and correspondingly stronger pinning in the SC phase due to the increased DW line tension. **g,** $I_{ac} = 0.4$ nA as in (a), showing a clear domain expansion and contraction (blue) near contact 4, demonstrating DW dynamics at sub-nA currents, the lowest reported driving current to date. **h,** $I_{ac} = 1.63$ nA, exhibiting substantially larger DW excursions than in (c). **i,** $I_{ac} = 4.91$ nA, showing a strongly enhanced DW response compared to (e).

Upon cooling to below $T_c$, the DW energy should acquire additional contribution from the Cooper-pair condensation energy. Since the pairing mechanism is unknown, for an order-of-magnitude estimate we consider a conventional BCS superconductor, for which the condensation energy density is given by [65]

$$U_{SC} = \frac{1}{2} \nu_0 \Delta^2,$$

where $\Delta \approx 1.76k_BT_c$ is the superconducting gap and $\nu_0$ is the density of states at the Fermi level. Since the DW line tension is set by the condensation energy density integrated over the wall width, $U_{\mathrm{SC}}$ provides an estimate of the additional pinning force $f_{\mathrm{pin}}$ arising from the suppression of the SC order parameter at the DW. Taking $T_c \approx$ 150 mK and an effective mass $m^* \approx 3.5m_e$ [14], and approximating the density of states as $\nu_0 = \frac{m^*}{2\pi\hbar^2}$, we obtain a bound $f_{\mathrm{pin}} \lesssim 3\times10^{-16}$ N/µm. Compared to our estimate of $f_{\mathrm{DW}} \approx 1.3\times10^{-15}$ N/µm, the rough nature of these estimates suggests that this additional contribution may account for the observed differences in hysteretic behavior above and below $T_c$.

In the presence of current, both the Zeeman energy gradient and the current-induced force act on the DW, and the coercive field is therefore expected to decrease with increasing current, as observed in Extended Data Fig. 4. The characteristic currents required to suppress $B_{co}$—of order a few tens of nA—are consistent with the force estimates above, further reinforcing this picture.

To benchmark the experimentally derived $f_{\mathrm{pin}}$, we use a previously estimated DW energy $U_{\mathrm{DW}} \approx 0.05$ meV/nm, derived for gapped systems [42,59,66,67], and a characteristic DW width $w$ of a few nm, yielding $f \sim U_{\mathrm{DW}}/w \sim 10^{-12}$ N/µm, i.e., two to three orders of magnitude larger than the measured value. Since in Chern insulators the DW energy is set by the bulk gap [59,66], its absence in the metallic 1/4M state is expected to substantially reduce the effective line tension.

**Dependence of DW dynamics on phase diagram and ramp rate**

To gain further insight into DW dynamics, we measured the dependence of the spike-like hysteretic transport on the magnetic-field sweep rate at several $n$ and $D$ values below the $n_{DW}(D)$ line at $T = 20$ mK (Extended Data Fig. 3). While the detailed features depend on specific parameters and exhibit stochastic fluctuations, several systematic trends emerge.

**1) Dependence on phase diagram**. The stability range on the HR state decreases monotonically with decreasing $|D|$. It spans the widest field range in the high-$D$ 1/4M phase (Extended Data Figs. 3a-d) and becomes extremely narrow in the low-$D$ CSC phase. In addition, the peak values of $R_{xx}$ and $|R_{xy}|$ increase gradually from a few kΩ in the high-$D$ 1/4M to about 25 kΩ in low-$D$ CSC. Understanding the origin of these trends requires further investigation.

**2) ZR state.** We find that the ZR state in the CSC phase (Figs. 1g–k) can transform into an LR state after extensive magnetic-field sweeps. Figures 2e,f,i,j, Fig. 3, and Extended Data Figs. 1d–k, 2i–k, 2m–o, and 3e–t illustrate such LR states. This ZR-to-LR transformation can also occur stochastically, as shown in Extended Data Figs. 3v,w. The ZR state can be restored by heating the sample above 10 K followed by zero-field cooling.

**3) Low-field LR state**. Initializing the system at negative $B$ selects the $K'\downarrow$ polarization as the majority isospin state. As the field reverses to positive, the opposite isospin becomes energetically favorable, and small minority $K\uparrow$ domains nucleate, likely at sample edges where intervalley mixing and disorder are strongest [68]. Additional small bubbles can also nucleate in the bulk due to fluctuations. At this stage, the minority domains remain spatially confined and do not span multiple contacts, consistent with the LR state observed in transport. Although the $K\uparrow$ polarization is energetically favorable, the expansion of these domains into the bulk is limited by DW pinning, and they therefore remain metastable until a larger field drives their growth, transforming the system into the HR state. Since the expansion of a DW can be viewed as the motion of an elastic string in a disorder-induced potential landscape, this process is slow and stochastic. The probability for a domain to expand during a field ramp increases with the duration of the ramp. Consequently, slower sweep rates allow more time for domain growth and yield lower characteristic fields for the LR-to-HR transition, whereas faster ramps suppress expansion and shift the transition to higher fields. This average trend is observed in Extended Data Fig. 3.

**4) HR state.** The HR state forms once the minority domains expand to sizes comparable to the device dimensions, effectively spanning multiple contacts and introducing highly resistive DWs across current paths. Further expansion, although energetically favorable, is again limited by the same stochastic process of overcoming disorder-induced pinning. As a result, the HR-to-LR transition field, at which the $K\uparrow$ domains expand to occupy most of the sample and become the majority phase, exhibits a similar ramp-rate dependence, increasing on average with faster ramps (Extended Data Fig. 3).

**5) High-field LR state**. At higher $B$, the $K\uparrow$ phase expands to form the new majority phase, while the residual $K'\downarrow$ domains shrink to small, isolated pockets, restoring the LR state with reversed minority polarization. Upon ramping the field down, the minority $K'\downarrow$ domains remain energetically unfavorable for $B > 0$, preserving the LR state and preventing the formation of an HR state. Only for $B < 0$ do they expand again, reestablishing the HR state and completing the hysteresis cycle (Extended Data Fig. 3).

**6) Hysteresis in the LR state.** Since the field ramp acts as an effective driving force—analogous to a *dc* current—a small hysteresis is expected even within the LR state, in addition to the large hysteresis characteristic of the HR state. In this regime, the minority domains respond to the effective force but remain confined. At low sweep rates, this hysteresis is barely discernible, but becomes clearly visible at higher ramp rates (see zoomed-in panels in Extended Data Fig. 3). This weak hysteresis underlies the small background signal observed in the CSC phase in Figs. 2i,j and in Extended Data Figs. 1d-k, 2i-k, and 2m-o.

**7) Possible heating effects.** Ramping the magnetic field can lead to sample heating. Indeed, at the highest ramp rates, we observe an increase of the base temperature by about 10 mK, as measured by a thermometer located near the sample. Such heating can increase the measured $R_{xx}$ and $R_{xy}$. However, this contribution is negligible compared to dominant effects of DW dynamics, for several reasons: (i) Increasing the temperature to 30 mK produces only a negligible change in resistance (see the lowest two curves in Fig. 1g corresponding to 20 and 60 mK). (ii) Since the heating power scales as $(dB/dt)^2$, it is symmetric with respect to sweep direction and cannot account for the observed hysteresis. (iii) In our ramp sequence, the ramp direction is reversed instantaneously at the maximal field, so the temperature should remain continuous. In contrast, both $R_{xx}$ and $R_{xy}$ exhibit abrupt changes at the field extrema. (iv) Heating would lead to similar changes in $R_{xx}$ and $R_{xy}$ for both sweep directions. Instead, $R_{xx}$ increases predominantly during the up-sweep, while the hysteresis in $R_{xy}$ exhibits a non-monotonic field dependence. (v) Finally, heating proportional to $(dB/dt)^2$ is independent of the instantaneous value of $B$. In contrast, the hysteresis loops in the LR state exhibit a sharp reversal of their sense—from clockwise to anticlockwise—precisely at $B = 0$, where the minority domains switch from energetically unfavorable to favorable (right two columns in Extended data Fig. 3). Taken together, these observations provide strong evidence that the hysteretic behavior in the LR state is governed by DW dynamics.

**8) Stochastic fluctuations.** The isospin domains exhibit pronounced random reconfigurations, giving rise to large resistance fluctuations, as directly visualized in Extended Data Fig. 5 alongside simultaneous transport measurements. DW motion leads to stochastic switching and resistance fluctuations that persist even under fixed external conditions. Extended Data Fig. 3v shows an example of large noise in $R_{xx}(t)$, measured continuously over several hours within the CSC phase under fixed conditions revealing large stochastic fluctuations. The zoomed-in view in Extended Data Fig. 3w, reveals a ZR state over the first ~20 min, interrupted by random short spikes reaching up to 3 kΩ. The system then undergoes a spontaneous transition to a LR state, where it remains for the rest of the measurement (~3 hours). Interestingly, in this LR state, sharp stochastic spikes persist, but they predominantly reduce the instantaneous resistance rather than increase it. Once small domains become trapped, the LR state is long-lived, but can be partially annealed by thermal cycling (green and brown curves in Figs. 2e,f and 3a,b). These pronounced fluctuations highlight the inherently stochastic nature of the multi-domain CSC phase, whose statistical properties warrant further investigation.

**Negative $R_{xx}$ state**

Previous studies have reported the surprising emergence of negative $R_{xx}$ during magnetic hysteresis measurements [8,14], but its origin has not been addressed. Extended Data Fig. 9a shows such a case in the HR state within the CSC phase. Upon initialization at negative (positive) $B$ and sweeping the field up (down), pronounced negative $R_{xx}$ states are observed alongside positive spikes in the HR state (blue and red curves). This configuration can be reproduced multiple times; however, each set of sweeps follows a somewhat different trajectory (green and brown curves), reflecting the inherent stochasticity of the HR state.

To investigate the origin of this phenomenon, we attempted to acquire a $B_z^{ac}$ map. However, once the magnetic-field sweep is stopped, the negative $R_{xx}$ state rapidly collapses before an image can be acquired. A more stable negative $R_{xx}$ state can instead be realized by applying a small *dc* current $I_{dc}$. We first initialize the system at $B = 120$ mT and sweeping the field to $B = -36$ mT, establishing an HR state with $R_{xx} \cong 1.7$ kΩ at $I_{dc} = 0$ nA. Applying $I_{dc} = -3.2$ nA drives the system into in a negative $R_{xx} \cong -400\ \Omega$ state. The switching between the high positive and negative $R_{xx}$ states can be reproducibly cycled, as shown in Extended Data Figs. 9c,d.

Extended Data Figs. 9l shows the $B_z^{ac}$ map acquired in the negative $R_{xx}$ state at $I_{dc} = -3.2$ nA and $B = -36$ mT, revealing the isospin domain dynamics in this state, which can be understood schematically as follows. Upon preparing the initial HR state at $I_{dc} = 0$ nA, the minority domain $K'\downarrow$ encompasses contacts 3 and 4 (Extended Data Fig. 9g). This configuration yields a large positive $R_{xx}$, as a highly resistive DW resides between contacts 2 and 3, analogous to the configuration discussed in Fig. 5j, and corresponding to the black markers in the simplified schematic $I$–$V$ plot in Extended Data Figs. 9e,f.

Applying a negative $I_{dc}$ expands the minority domain via electron injection at contact 4, resulting in the configuration shown in Extended Data Fig. 9h, corresponding to the pink markers in Extended Data Figs. 9e,f. During the positive half-cycle of $I_{ac}$, the reduced electron injection causes the minority bubble to contract slightly as illustrated in Extended Data Fig. 9i (orange dot in Extended Data Fig. 9e), whereas during the negative half-cycle, enhanced injection leads to a slight expansion illustrated in Extended Data Fig. 9j (green dot in Extended Data Fig. 9e). Extended Data Fig. 9k shows the difference between these two configurations, reproducing the measured $B_z^{ac}$ map in Extended Data Fig. 9l. The negative $R_{xx}$ in this configuration arises from a negative differential response $dV_{xx}/dI$ (pink dot in Extended Data Fig. 9f), corresponding to the negative difference between the orange and green dots in Extended Data Fig. 9e. The schematic $I$-$V$ characteristics in Extended Data Figs. 9e,f further suggest that negative $R_{xx}$ should also occur at positive $I_{dc}$; however, this configuration appears to be less stable and is difficult to capture experimentally.

Notably, while negative differential resistance at finite bias is common in electronic amplifiers, its occurrence at zero bias would appear to violate local thermodynamic equilibrium. Consistently, at $I_{dc} = 0$ nA, $R_{xx}$ remains positive (Extended Data Figs. 9d,f). However, during magnetic-field sweeps, $R_{xx}$ is found to become negative even at nominally zero $I_{dc}$ (Extended Data Fig. 9a). We therefore attribute the emergence of negative $R_{xx}$ in this regime to an effective bias generated by the ramping magnetic field, equivalent to a finite *dc* current.

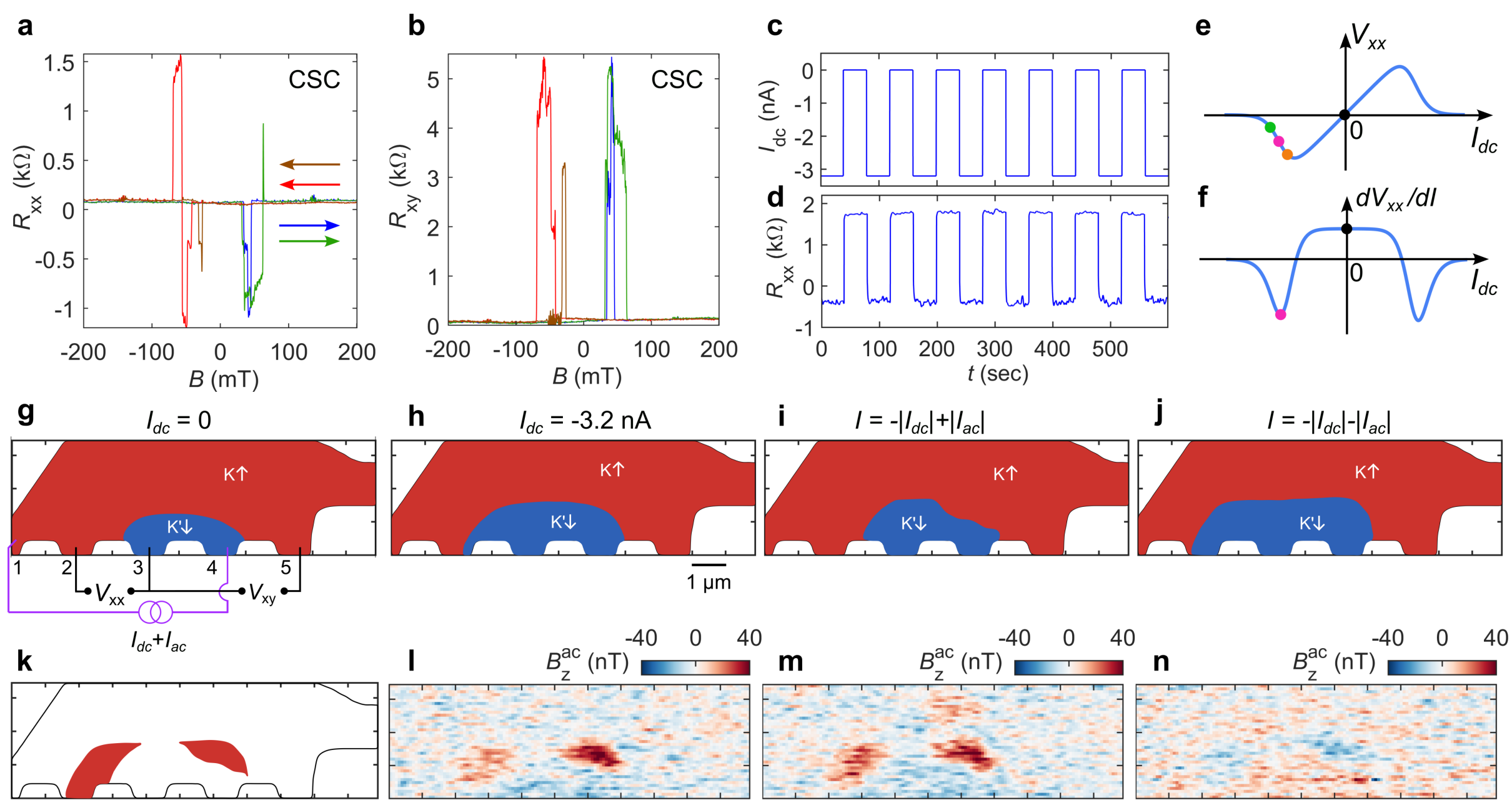

**Extended Data Fig. 9. Isospin domain configuration in the negative $R_{xx}$ state in the CSC phase. a,** Magnetic hysteresis of $R_{xx}$ measured in the CSC phase in device D1 ($D = -0.939$ V/nm, $n = 0.58\times10^{12}$ cm$^{-2}$, $T = 20$ mK), showing pronounced negative $R_{xx}$ in the HR state using a measurement current $I_{ac} = 1$ nA in the absence of *dc* current ($I_{dc} = 0$ nA). Two independent sets of field-sweeps—up (blue, green) and down (red, brown)—are shown, highlighting the stochastic component of the domain configurations in the HR state, as well as the small but finite resistance in the LR state. **b,** Corresponding hysteresis of $R_{xy}$. **c,** Time traces of the *dc* current $I_{dc}$ applied between contacts 4 and 1, switching between 0 and $-3.2$ nA. **d,** Corresponding $R_{xx}$ measured using $I_{ac} = 1.5$ nA at $B = -36$ mT following initialization at $B = 120$ mT, showing $R_{xx} \cong 1.7$ kΩ at $I_{dc} = 0$ nA and negative $R_{xx} \cong -400$ Ω at $I_{dc} = -3.2$ nA. **e,** Schematic $I$-$V$ characteristics reflecting the *dc* potential difference $V_{xx}$ between contacts 3 and 2 arising from current-induced reconfiguration of the isospin domains by $I_{dc}$. **f,** Corresponding schematic differential resistance $dV_{xx}/dI$ vs. $I_{dc}$. **g,** Schematic isospin domain map illustrating spontaneous formation of a minority $K'$↓ bubble by magnetic hysteresis at $I_{dc} = 0$ nA. The minority domain encompasses contacts 3 and 4, resulting in a large $R_{xx}$ due to DW between contacts 2 and 3, indicated by black dot in (e,f). **h,** Schematic of the expanded minority bubble due to electron injection at contact 4 for $I_{dc} = -3.2$ nA, corresponding to the pink dot in (e,f). **i,** During the positive half-cycle of $I_{ac}$, the minority domain contracts due to reduced electron injection at contact 4 for negative $I_{dc} = -3.2$ nA. The corresponding negative $V_{xx}$ increases (orange dot in (e)), relative to (h), as contacts 2 and 3 become fully separated into majority and minority domains. **j,** During the negative half-cycle of $I_{ac}$, the minority domain expands due to enhanced electron injection at contact 4. The corresponding negative $V_{xx}$ decreases (green dot in (e)) relative to (h), as partial equilibration occurs between the potentials of contacts 2 and 3 within the minority domain. **k,** Difference between (i) and (j), reproducing the measured $B_z^{ac}$ map in (l). The resulting $R_{xx} = dV_{xx}/dI$, determined by the difference between the orange and green dots in (e), is negative, as indicated by the pink dot in (f). **l,** $B_z^{ac}$ map measured at $I_{dc} = -3.2$ nA and $B = -36$ mT, revealing the isospin domain dynamics in the negative $R_{xx}$ state consistent with the schematic in (k). **m,** Another example of negative $R_{xx}$ state similar to (l), measured at $I_{dc} = -3.5$ nA and $B = -36$ mT. **n,** $B_z^{ac}$ map measured at $I_{dc} = -3.0$ nA and $B = 27$ mT, following initialization at $B = -120$ mT, showing additional negative $R_{xx}$ state similar to (l), but with opposite signal polarity due to the reversed $B$.

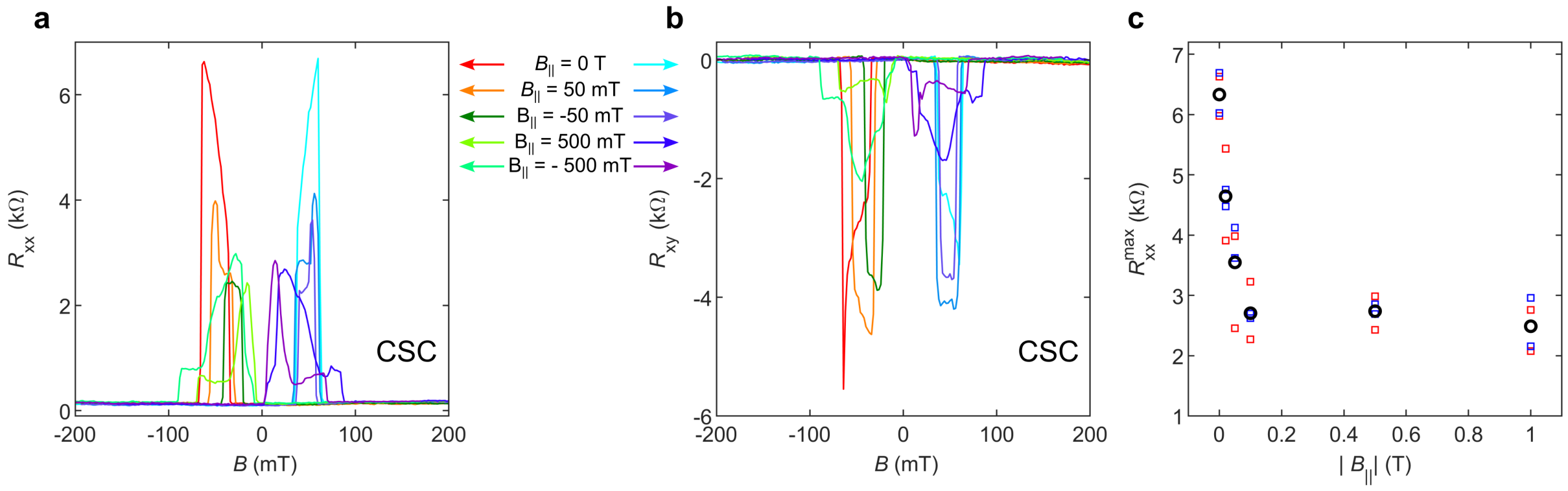

**Extended Data Fig. 10. In-plane magnetic-field dependence of transport hysteresis in the CSC phase.** **a,b,** Magnetic-field hysteresis of $R_{xx}$ (**a**) and $R_{xy}$ (**b**) measured in the CSC phase in device D1 ($D = -0.968$ V/nm, $n = 0.63\times10^{12}$ cm$^{-2}$, $T = 20$ mK) while sweeping the out-of-plane field $B$ up and down (ramp rate $|dB/dt| = 50$ mT/min), in the presence of fixed indicated in-plane fields $B_{\parallel}$. **c,** Peak resistance $R_{xx}^{max}$ vs. $B_{\parallel}$, showing suppression of $R_{xx}$ in the HR state with increasing in-plane field. For each $B_{\parallel}$, the black circles denote $R_{xx}^{max}$ averaged over four peak values (open squares) at positive and negative $B$ for sweep-up (blue) and sweep-down (red) traces. At $B_{\parallel} = 0$ T, electron transmission across the DW separating opposite chiralities is constrained by both intervalley momentum mismatch and spin conservation. Applying $B_{\parallel}$ tilts the spins, partially reducing the spin mismatch and thereby lowering the DW resistivity $\rho_{DW}$, which saturates at a value dominated by valley momentum mismatch.

### Suppression of DW energy by a hidden order

We consider a phenomenological 2D Landau free-energy functional, meant to describe the coexistence of an isospin-polarization order parameter $\phi$ and a secondary, isospin-independent order parameter $\Psi$,

$$F = F_{\text{IP}} + F_{\text{sec}} + F_{int}.$$

The isospin-polarization sector can be described by

$$F_{\text{IP}} = g\int d^2\boldsymbol{x}\left[\xi_\phi^2(\nabla\phi)^2 + (\phi-1)^2(\phi+1)^2 + b(\boldsymbol{x})\phi\right].$$

Here, $g$ is the condensation energy density of the isospin order, $\xi_\phi$ is the associated correlation length, and $b(\boldsymbol{x})$ is a disorder field which couples linearly to $\phi$. The order parameter $\phi = \pm1$ represents polarization into one of the isospin flavors.

We now consider a secondary intra-isospin order parameter $\Psi$. Whereas the isospin-polarization transition can be traced to electronic interactions between electrons of different flavors, this order can emerge due to minimization of the intra-flavor interactions. Thus, one may assign an energy gain of $\delta E$ to this order, per isospin flavor where $\Psi$ condenses.

Intuitively, if the system is in a parameter regime where it can lower its energy by triggering an intra-isospin reconstruction, more energy could be gained if this reconstruction is duplicated in other flavors. Consequently, in the "$\phi$-paramagnet" phase, where two flavors are occupied and $\phi = 0$, the system can gain more energy by condensing $\Psi$, as compared to the isospin-polarized $\phi = \pm1$ phases.

The above considerations are succinctly captured by the following,

$$F_{\text{sec}} = f\int d^2\boldsymbol{x}\left[\xi_\Psi^2|\nabla\Psi|^2 + \frac{r}{2}|\Psi|^2 + \frac{1}{4}|\Psi|^4\right],$$

$$F_{\text{int}} = \lambda\int d^2\boldsymbol{x}(\phi^2-1)|\Psi|^2.$$

In the isospin-polarized phase $F_{\text{int}}$ vanishes, and the secondary order develops for $r < 0$, with condensation energy density $\frac{1}{4} f r^2$, and correlation length $\xi_\Psi$. Conversely, in a region where the system is $\phi$-paramagnetic ($\phi = 0$), $F_{\text{int}}$ lowers the overall energy (for $\lambda > 0$). Consequently, one may consider an effective potential for the isospin-polarization order parameter by integrating out the uniform $\Psi$, $V(\phi, r) = (\phi^2 - 1)^2 + \frac{\lambda |r| \Theta(-r)}{g} (\phi^2 - 1)$.

We now estimate the line tension $J_{\text{DW}}$ of a DW separating oppositely isospin-polarized domains,

$$J_{\text{DW}}(r) \approx 2g\xi_\phi \int_0^{\phi_{\min}} d\phi \sqrt{V(\phi, r) - V(\phi_{\min}, r)} = \frac{4}{3} g \xi_\phi \left(1 - \frac{\lambda |r| \Theta(-r)}{2g}\right)^{\frac{3}{2}},$$

where $\phi_{\min}$ is the location of the minimum of this effective potential. Clearly, the coupling to the secondary order can substantially lower the line tension (Extended Data Fig. 11), making the system more susceptible to appearance of domains driven by intrinsic disorder fields $b(\boldsymbol{x})$.

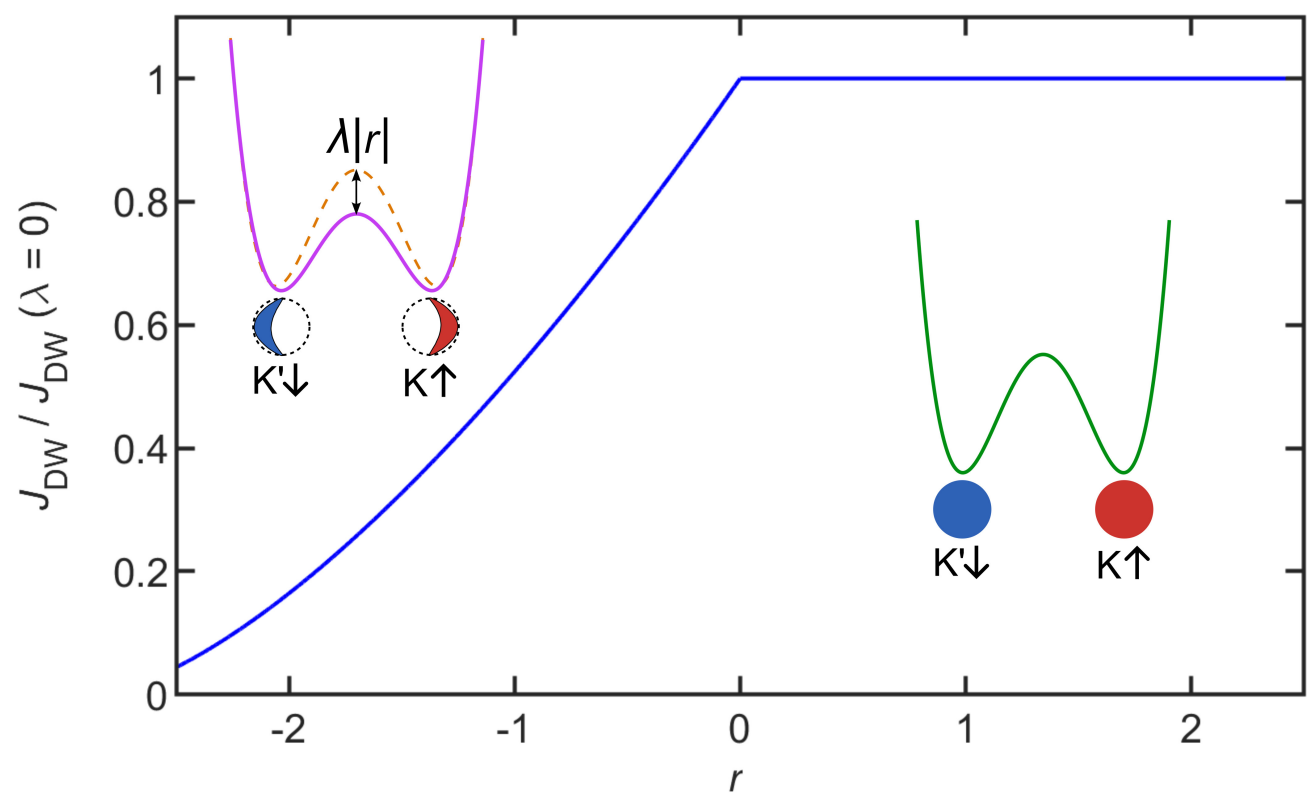

**Extended Data Fig. 11. DW line-tension suppression by a hidden order.** Calculated DW line tension $J_{\text{DW}}$ of the isospin-polarization order parameter, as a function of $r$, which tunes the transition of the hidden secondary order. $J_{\text{DW}}$ is normalized to its value in the decoupled limit ($\lambda = 0$). For $r > 0$, the secondary order is absent ($|\Psi| = 0$). For $r < 0$, the secondary order condenses, $|\Psi| \approx \sqrt{-r}$, lowering the free-energy barrier between the degenerate isospin-polarized states by $\approx \lambda |r|$ (left inset), and thereby suppressing the DW line tension. In this plot, we use $\lambda = 0.7g$.

One possibility for this hidden order is breaking the $C_3$ symmetry intrinsic to rhombohedral graphene, distorting the shape of the Fermi surfaces (or selectively populating a subset of the three valley Fermi pockets). Notice $F_{\text{sec}}$ is written as a $U(1)$ symmetry breaking theory, which does not affect our conclusions. To take the $C_3$ symmetry into account, one can include an additional term

$$F_{\text{sec}} \to F_{\text{sec}} + f \int d^2\boldsymbol{x} \frac{2w}{3} (\Psi^3 + \Psi^{*3}).$$

This type of symmetry breaking has been observed experimentally in bilayer and rhombohedral trilayer graphene [15,69–71], and has recently been linked to the observation of an in-plane anomalous Hall response in RMG [72]. Notably, the onset of this symmetry-breaking state appears to correlate with SC in multilayer graphene, and a Cooper pairing mechanism linking the symmetry-broken state to the SC phase has been proposed [73]. Within this framework, the emergence of CSC naturally coincides with a reduction in DW energy.

We emphasize, however, that alternative scenarios, e.g. involving a Lifshitz-like transition of the single-particle band within the 1/4M phase (see [24]), cannot be excluded. Establishing a direct connection between such a transition and the reduction of DW energy on the one hand, and the onset of CSC on the other, will require further investigation.

| Figure | Initialization field $B_{int}$ (mT) | Measurement field $B$ (mT) | Gate or current *ac* excitation | Scan height (nm) | $T$ (mK) |
|---|---|---|---|---|---|
| Fig. 1h | 150 | 26 | $V_{bg}^{ac}(SW) = 0.1$ Vrms, $f = 1.193$ kHz | 140 | 20 |
| Fig. 1i | -150 | -28 | $V_{bg}^{ac}(SW) = 0.1$ Vrms, $f = 1.193$ kHz | 140 | 20 |
| Fig. 1j | 120 | 20 | $V_{bg}^{ac}(SW) = 0.1$ Vrms, $f = 1.193$ kHz | 140 | 20 |
| Fig. 1k | -100 | 20 | $V_{bg}^{ac}(SW) = 0.1$ Vrms, $f = 1.193$ kHz | 140 | 20 |
| Fig. 1m | 250 | 16 | $V_{tg}^{ac}(SqW) = 0.35$ Vpp, $f = 1.181$ kHz | 200 | 20 |
| Fig. 1o | −250 | 25.5 | $V_{tg}^{ac}(SqW) = 0.35$ Vpp, $f = 1.181$ kHz | 200 | 20 |
| Fig. 2o | −200 | 35 | $V_{bg}^{ac}(SW) = 0.2$ Vrms, $f = 1.173$ kHz | 150 | 20 |
| Fig. 2p | −200 | 13 | $V_{bg}^{ac}(SW) = 0.2$ Vrms, $f = 1.173$ kHz | 150 | 20 |
| Fig. 3c | -200 | 15 | $V_{bg}^{ac}(SW) = 0.15$ Vrms, $f = 1.173$ kHz | 180 | 20 |
| Fig. 3d | 200 | 10 | $V_{bg}^{ac}(SW) = 0.15$ Vrms, $f = 1.173$ kHz | 180 | 20 |
| Fig. 3g | -200 | 14 | $V_{bg}^{ac}(SW) = 0.15$ Vrms, $f = 1.173$ kHz | 180 | 20 |
| Fig. 3h | 200 | -19.5 | $V_{bg}^{ac}(SW) = 0.15$ Vrms, $f = 1.173$ kHz | 180 | 20 |
| Fig. 4a | −150 | 13 | $V_{bg}^{ac}(SqW) = 0.4$ Vpp, $f = 2.133$ kHz | 200 | 20 |
| Fig. 4b | −150 | 13 | $V_{bg}^{ac}(SqW) = 0.4$ Vpp, $f = 2.133$ kHz | 200 | 20 |
| Fig. 4c | −150 | 13 | $V_{bg}^{ac}(SqW) = 0.4$ Vpp, $f = 2.133$ kHz | 200 | 20 |
| Fig. 4d | −150 | 13 | $V_{bg}^{ac}(SqW) = 0.4$ Vpp, $f = 2.133$ kHz | 200 | 20 |
| Fig. 4e | −150 | 13 | $V_{bg}^{ac}(SqW) = 0.4$ Vpp, $f = 2.133$ kHz | 200 | 20 |
| Fig. 4f | −150 | 13 | $V_{bg}^{ac}(SqW) = 0.4$ Vpp, $f = 2.133$ kHz | 200 | 20 |
| Fig. 4i | 130 | −25 | $I_{ac}(SqW) = 5.3$ nArms, $f = 1.181$ kHz | 140 | 20 |
| Fig. 4m | 40 | −20 | $I_{ac}(SqW) = 6.5$ nArms, $f = 1.181$ kHz | 140 | 20 |
| Fig. 5d | −130 | 27 | $I_{ac}(SqW) = 2$ nArms, $f = 433.33$ Hz | 140 | 20 |
| Fig. 5h | −130 | 27 | $I_{ac}(SqW) = 2$ nArms, $f = 433.33$ Hz | 140 | 20 |
| ED Fig. 2d | - | 58 | $V_{bg}^{ac}(SqW) = 0.565$ Vpp, $f = 2.133$ kHz | 80 | 20 |
| ED Fig. 5a | 33 | −21 | $I_{ac}(SqW) = 10$ nArms, $f = 1.181$ kHz | 140 | 20 |
| ED Fig. 5d | −21 | 15 | $I_{ac}(SqW) = 10$ nArms, $f = 1.181$ kHz | 140 | 20 |
| ED Fig. 5g | 100 | −22 | $I_{ac}(SqW) = 7$ nArms, $f = 1.181$ kHz | 140 | 550 |
| ED Fig. 5j | −20 | 20 | $I_{ac}(SqW) = 11.5$ nArms, $f = 1.181$ kHz | 140 | 550 |
| ED Fig. 6a-d | −150 | 13 | $V_{bg}^{ac}(SqW) = 0.4$ Vpp, $f = 2.133$ kHz | 200 | 20 |
| ED Fig. 9l | 130 | −36 | $I_{ac}(SqW) = 1.5$ nArms, $f = 433.33$ Hz | 140 | 20 |
| ED Fig. 9m | 130 | −36 | $I_{ac}(SqW) = 1.95$ nArms, $f = 433.33$ Hz | 140 | 20 |
| ED Fig. 9n | −120 | 27 | $I_{ac}(SqW) = 1.63$ nArms, $f = 433.33$ Hz | 140 | 20 |

**Extended Data Table 1. Summary of SOT imaging parameters in the various figures.** $SqW$: square wave excitation. $SW$: sine wave excitation. In high-frequency current-modulation imaging, the tabulated $I_{ac}$ values represent the nominal *ac* current, determined from the rms *ac* voltage measured across a 12 kΩ resistor in series with the sample at the current source contact (contact 4). These nominal values are quoted in the figures. However, the actual current flowing through the sample is expected to be significantly lower due to the low-pass filters installed on the electrical lines, whose cutoff frequency is comparable to or below the drive frequency.